# Purcell Enhancement of Parametric Luminescence:
# Bright and Broadband Nonlinear Light Emission in Metamaterials

Artur R. Davoyan[1,†] and Harry A. Atwater[1,‡]
[1]Resnick Sustainability Institute and Kavli Nanoscience Institute,
California Institute of Technology
1200 E. California blvd., Pasadena, 91125, CA
[†]davoyan@caltech.edu
[‡]haa@caltech.edu

*Abstract:*
　　Single-photon and correlated two-photon sources are important elements for optical information systems. Nonlinear downconversion light sources are robust and stable emitters of single photons and entangled photon pairs. However, the rate of downconverted light emission, dictated by the properties of low-symmetry nonlinear crystals, is typically very small, leading to significant constrains in device design and integration. In this paper, we show that the principles for spontaneous emission control (i.e. Purcell effect) of isolated emitters in nanoscale structures, such as metamaterials, can be generalized to describe the enhancement of nonlinear light generation processes such as parametric down conversion. We develop a novel theoretical framework for quantum nonlinear emission in a general anisotropic, dispersive and lossy media. We further find that spontaneous parametric downconversion in media with hyperbolic dispersion is broadband and phase-mismatch-free. We predict a 1000-fold enhancement of the downconverted emission rate with up to $10^5$ photon pairs per second in experimentally realistic nanostructures. Our theoretical formalism and approach to Purcell enhancement of nonlinear optical processes, provides a framework for description of quantum nonlinear optical phenomena in complex nanophotonic structures.

*Main Text:*
　　The Purcell effect is an elegant manifestation of quantum engineering by which the spontaneous emission rates of quantum emitters can be dramatically altered [Figs. 1 (a) and 1(b)] to tailor the design of nanoscopic and integrated single photon sources [1-3]. Nonetheless, the complexity of emitter design, and challenges associated with matching photon sources in frequency, polarization and phase have to date limited the use of individual quantum emitters in quantum optical systems. Nonlinear optical processes (e.g., spontaneous parametric downconversion [4-6] and four wave mixing [7]) offer a distinctly different approach to light generation. Their relative simplicity, high single photon indistinguishability, stability, and straightforward room-temperature entanglement make quantum nonlinear sources advantageous for a large variety of practical applications [8, 9], as well as in benchmark quantum experiments [10]. However sources based on quantum nonlinear processes suffer from a number of limitations, including phase mismatch, which constrains operation to a narrow, material-specific frequency band, and low efficiency (i.e., low photon pair generation rate per unit length), leading to bulky devices that do not lend themselves to compact monolithic integration [2]. We note that recent reports of waveguide-integrated structures can achieve high efficiency [7, 11-14], but are typically at least 100 microns long or utilize resonant cavities with very high quality factors [5, 6, 15, 16]. Mitigating these constraints would enable high efficiency quantum nonlinear sources of single and entangled photons for chip-scale optical devices.

Here we show that by modifying light dispersion and the density of optical states in hyperbolic metamaterials, photon pair generation through spontaneous parametric downconversion may be enhanced over a broad frequency range. We develop a comprehensive theoretical formalism describing quantum nonlinear light emission in structures with a modified density of optical states, such as nonlinear metamaterials, highly dispersive crystals, and plasmonic nanostructures. We further identify new regimes of nonlinear light generation, including phase mismatch free, wavelength tunable and hyperbolic photon pair emission. Finally, we discuss experimental feasibility and design.

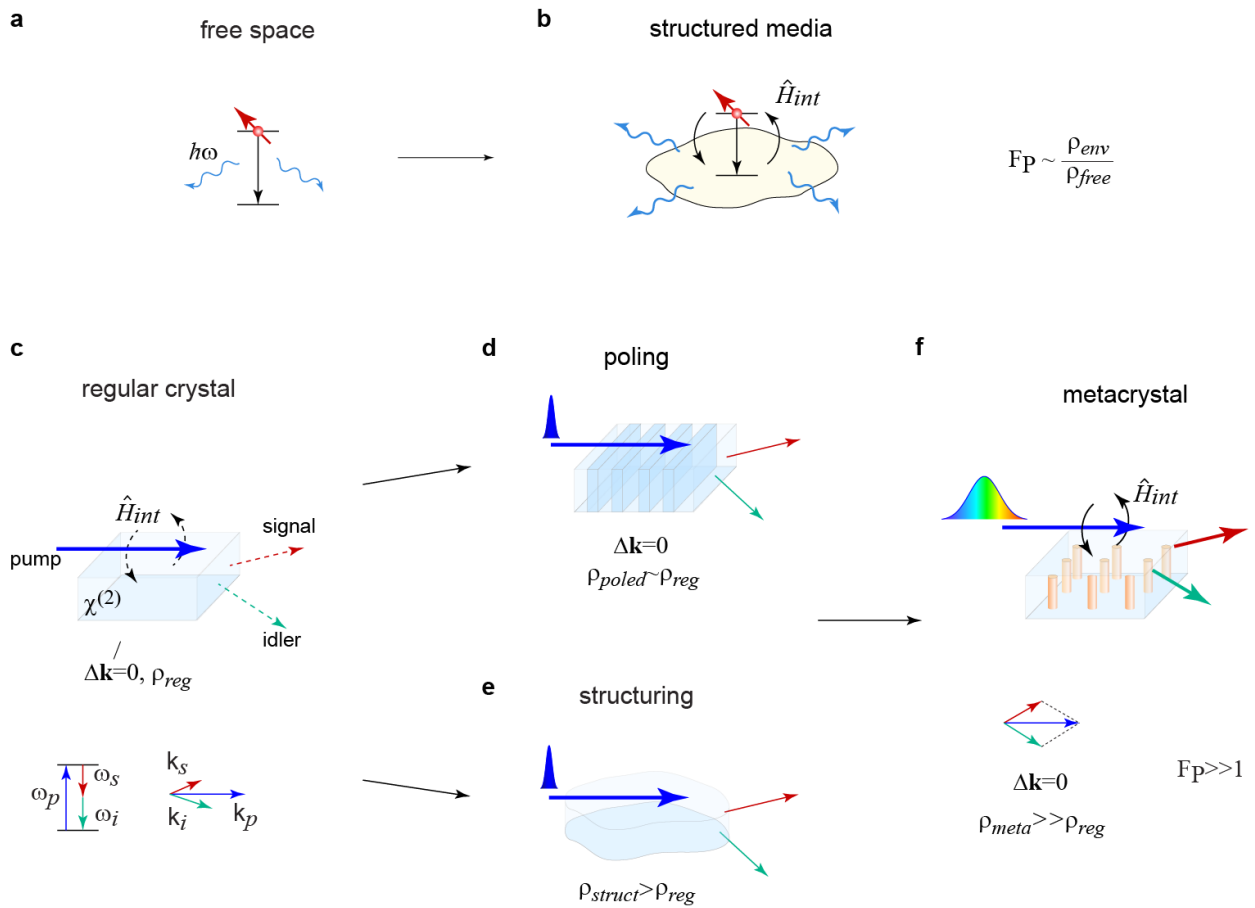

*FIG. 1 Purcell enhancement of nonlinear generation with metamaterials. (a) An excited two-level system in free space decays by spontaneous emission. This process may be enhanced by modifying the emitter coupling to the photonic environment, (b). (c) Light may also be spontaneously emitted within a nonlinear crystal, when pump photons spontaneously fission creating quantum-correlated signal-idler photon pairs. Nonlinear generation depends on the density of optical states, $\rho_{reg}$, and the strength of light-material interaction $\hat{H}_{int}$, which also depends on phase matching between the pump, signal and idler waves ($\Delta \boldsymbol{k}$) – a condition hard to meet in regular materials. (d) Poling of a nonlinear crystal minimizes the phase mismatch ($\Delta \boldsymbol{k} \to 0$) for a spectrally narrow operation range, but in general the density of optical modes is not significantly changed ($\rho_{poled} \simeq \rho_{reg}$). (e) Modification of the density of optical states in a resonator or waveguide enhances emission and modifies the phase-matching conditions, but the high quality factor limits the operational frequency range. (f) Conversely, metamaterials may enable nonresonant, broadband, phase-mismatch free Purcell enhancement of spontaneous nonlinear light emission.*

In spontaneous parametric downconversion, pump photons with frequency $\omega_p$ in a quadratically nonlinear crystal may spontaneously fission [Fig. 1(c)] to emit quantum-correlated signal and idler photons with frequencies $\omega_s$ and $\omega_i$, respectively, and wavevectors $\boldsymbol{k}_s(\omega_s)$ and $\boldsymbol{k}_i(\omega_i)$, where energy conservation requires that $\omega_p = \omega_i + \omega_s$. The rate at which individual downconverted photons or photon pairs are generated is a key performance metric.

As is the case for ordinary spontaneous emission [Figs. 1(a) and 1(b)], spontaneous nonlinear luminescence depends on the strength of the quantum mechanical interaction and the density of available optical states in the system, $\rho \propto \int_{\partial \mathcal{V}_k} \left|\frac{\partial \omega}{\partial \boldsymbol{k}}\right|^{-1} d^2\boldsymbol{s}$ (for an unbounded medium, where integration is over the isofrequency surface $\partial \mathcal{V}_k$) [17]. However spontaneous parametric downconversion also requires phase matching between the interacting pump, signal, and idler waves, $\Delta \boldsymbol{k} = \boldsymbol{k}_p - \boldsymbol{k}_s - \boldsymbol{k}_i \to 0$, see Fig. 1(c). Describing the Purcell-like enhancement of nonlinear luminescence requires modifying both the density of optical modes, determined by the isofrequency surface, $\partial \mathcal{V}_k$, and the light dispersion, $\boldsymbol{k}(\omega)$, at the pump, signal, and idler wavelengths. We note that periodic poling of a nonlinear crystal helps to minimize the phase matching constraint, but does not by itself significantly alter the density of optical states in the bulk of a nonlinear crystal [Fig. 1(d)]. The density of states can be tailored by using high Q cavities and resonators [Fig. 1(e)], but this imposes a sensitive phase matching requirement that reduces the downconversion bandwidth. By contrast, metamaterials designed with tailored subwavelength nanoscale structures can exhibit effective electromagnetic properties not readily available in nature. Nanophotonic materials with unusual material parameters have previously demonstrated potential for nonlinear optical generation [18 - 22] and also for tuning radiation properties of isolated quantum emitters [23 - 25]. The possibility of modifying spontaneous four wave mixing in third order nonlinear hyperbolic metamaterials was considered in [26]. We develop here a general theoretical approach, different from previous works, for use of metamaterial and plasmonic structures to enhance and control quantum nonlinear optical processes, particularly, spontaneous parametric downconversion, Fig. 1.

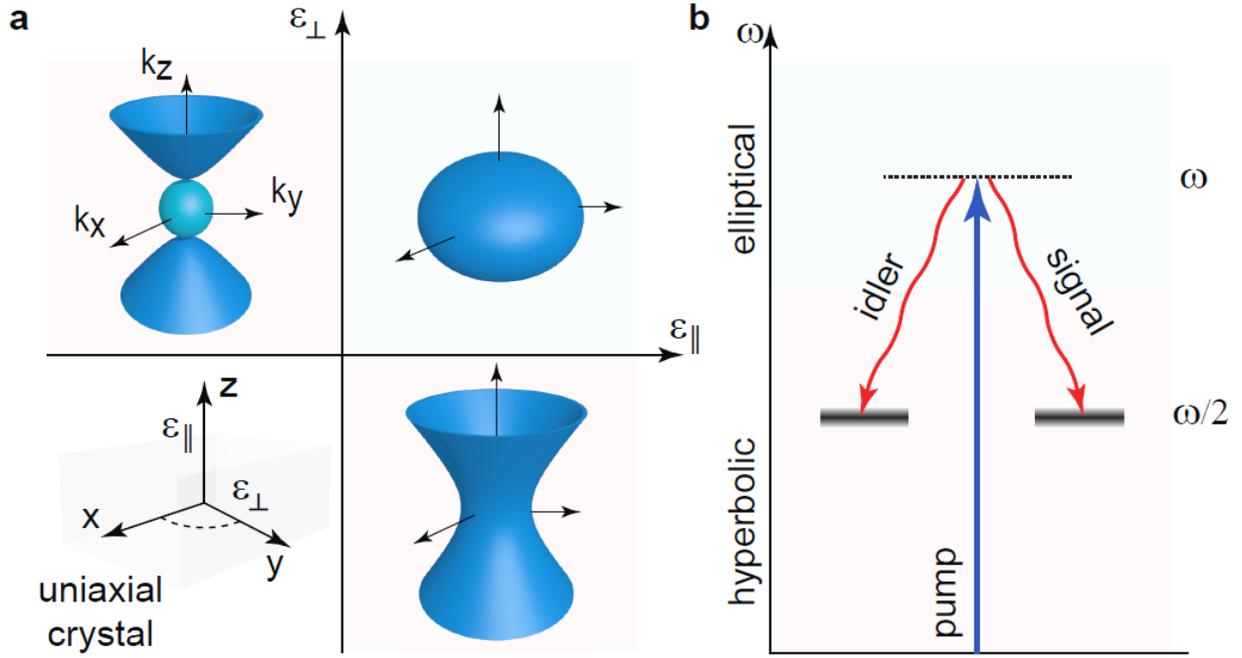

FIG. 2 Spontaneous parametric downconversion in hyperbolic metamaterials. (a) A map of possible isofrequency surfaces ($\partial \mathcal{V}_k$) for a uniaxial crystal. Elliptical and hyperbolic dispersion regimes may be accessed by controlling the signs of ordinary, $\varepsilon_\perp$, and extraordinary, $\varepsilon_\parallel$, permittivities. Notably, a semi-infinite number of optical modes are available in materials with hyperbolic dispersion ($\rho \propto \partial \mathcal{V}_k$). (b) Energy diagram of the downconversion process treated here. The metamaterial is pumped in an elliptical dispersion regime, whereas spontaneous downconversion of the signal-idler pairs occurs in the hyperbolic dispersion regime.

To illustrate the principles for modifying spontaneous downconversion with metamaterials, we consider a uniaxial crystal with an effective tensor permittivity $\bar{\bar{\varepsilon}} = \text{diag}(\varepsilon_\perp, \varepsilon_\perp, \varepsilon_\parallel)$. Figure 2(a) shows the isofrequency surfaces for different variations of the ordinary ($\varepsilon_\perp$) and extraordinary ($\varepsilon_\parallel$) components of the permittivity tensor. For regular crystals ($\varepsilon_\parallel > 0$ and $\varepsilon_\perp > 0$), the isofrequency surfaces are ellipsoidal [Fig. 2(a)]. The closed topology of these surfaces implies that the density optical modes is finite (i.e., $\int_{\partial \mathcal{V}_k} \left|\frac{\partial \omega}{\partial k}\right|^{-1} d^2 s \ll \infty$). Phase matching occurs only for a certain finite range of pump wavevectors [Fig. 3(a)].

In metamaterials, a different regime can be accessed, where either $\varepsilon_\parallel < 0$ or $\varepsilon_\perp < 0$, [23, 27, 28]. This is possible, for instance, in alternating subwavelength metal-dielectric structures, as shown schematically in Fig. 3(d), or in bulk crystals with pronounced material resonances, e.g., hexagonal boron nitride and bismuth selenide [29]). In such materials, the isofrequency surfaces for extraordinary waves are transformed into hyperboloids [Fig. 2(a)] [23, 27, 28]. Hence a large optical mode density becomes accessible (ideally infinite, but in practice it is limited by losses and the achievable minimum period of the layered structure, i.e., $\max(|\boldsymbol{k}|) \simeq \frac{2\pi}{\Lambda}$). Open hyperbolic isofrequency dispersion surfaces, in contrast with regular dispersion surfaces, remove phase matching constraints – that is, for any pump wavevector there is always a pair of signal-idler photons such that $\Delta \boldsymbol{k} = 0$ (see Figs. 3(c), 3(d), and 3(f), and a corresponding discussion in

Appendix B and Appendix C). Phase mismatch-free operation in the hyperbolic regime, as we show below, enables broadband operation.

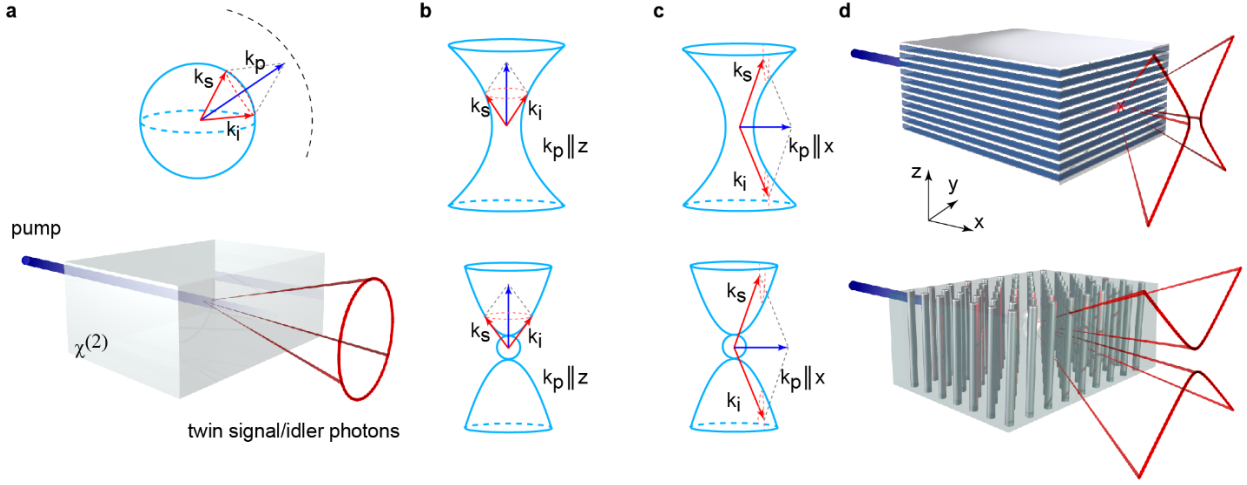

FIG. 3. Phase-mismatch-free spontaneous nonlinear light emission. (a) Phase matching diagram and corresponding photon emission pattern for an isotropic medium. For this closed isofrequency manifold, phase matching is possible only for a limited range of pump wavevectors ($|\mathbf{k}_p| \leq 2|\mathbf{k}_s|$). The dashed line denotes a condition beyond which phase matching is not possible. The photon emission is conical in this case. (b,c) Schematic of phase matching in layer- and wire-like hyperbolic metamaterials for pump propagation along the metamaterial axis. Open hyperbolic isofrequency surfaces for both of the hyperbolic cases imply pump wavevector independent phase-mismatch-free operation. The light emission is conical, similar to the isotropic case [panel (a)]. (d) A phase-mismatch free condition is also obtained for pump propagating along the $\mathbf{x}$ axis of the hyperbolic layer and wire metamaterials, respectively. However in this case signal and idler waves encompass an infinitely large state space enabling a significant Purcell-like enhancement of nonlinear luminescence. Panel (d) shows the expected emission patterns in red for this pump propagation configuration.

From a simple phase matching analysis, we can specify the preferred directions for single photon emission. Conical emission is anticipated for conventional crystals [Figs. 3(a)] and for hyperbolic metamaterials pumped along the crystal axis ($\mathbf{k}_p||\mathbf{z}$), see Fig. 3(b). However, for a pump perpendicular to the metamaterial axis ($\mathbf{k}_p||\mathbf{x}$) the expected light emission is hyperbolic [Figs. 3(c) and 3(d)]. We predict that it is this combination of phase-mismatch free parametric downconversion together with a near-infinite number of available optical modes that leads to a substantial enhancement of signal-idler downconverted photon pair emission rate. (Note that, a similar criterion is satisfied for some other spontaneous wavemixing processes, e.g., spontaneous four wavemixing [26]).

To probe the spontaneous parametric downconversion in the hyperbolic regime, we consider a process schematically shown in Fig. 2(b). We assume that photons of a pump laser beam in an elliptical dispersion regime spontaneously downconvert to extraordinary signal and extraordinary idler waves, both within a hyperbolic dispersion regime. To be specific, we consider a continuous plane-wave pump propagating along the $\mathbf{x}$ axis (as shown in Figs. 3(c) and 3(d)). We develop a general quantum mechanical model that predicts the downconversion rate in a range of complex structures, including such extremely anisotropic uniaxial metamaterials (see Appendix F for a detailed discussion). In contrast to previous works (e.g., Ref. [26]), our theory is based on the comprehensive eigen-mode analysis, which enables a deeper insight to key physical process in a variety of complex systems. We find that the emitted signal photon spectral power density may be estimated as:

$$\frac{dP_s}{d\lambda_s} = \frac{\hbar \pi c^3 L^2}{\lambda_s^4 \lambda_i} \frac{P_p}{\varepsilon_0 n_p} \int d^2 \boldsymbol{k}_{s\perp} c_{\boldsymbol{k}_s}^2 c_{\boldsymbol{k}_i}^2 \frac{\partial k_{s\parallel}}{\partial \omega} \frac{\partial k_{i\parallel}}{\partial \omega} N(\boldsymbol{k}_s) \left| \frac{1 - e^{i\Delta kL}}{i\Delta \boldsymbol{k} L} \right|^2 e^{-\gamma'(\boldsymbol{k}_s)L} \tag{1}$$

where $N(\boldsymbol{k}_s) = \left| \sum \bar{\bar{\chi}}^{(2)}_{lmn} u_l(\omega_p) u_m(\omega_s) u_n(\omega_i) \right|^2$ corresponds to a nonlinear media 'polarization mixing' term with $\bar{\bar{\chi}}^{(2)}$ being a nonlinear susceptibility and $\boldsymbol{u}_{p,s,i}$ corresponding to polarization of pump, signal or idler waves, $\lambda_{s,i}$ are signal and idler wavelengths, $P_p$ is the pump power in the crystal, $n_p$ is effective index at the pump wavelength, $c_{\boldsymbol{k}_{s,i}}$ are the coefficients due to quantization of the interacting fields, $d^2 \boldsymbol{k}_\perp = dk_{sy} dk_{sz}$ with $k_{sy}$ and $k_{sz}$ being $y$ and $z$ components of the signal wavevector $\boldsymbol{k}_s$, $\frac{\partial k_{s,i\parallel}}{\partial \omega}$ refer to group velocities of signal and idler waves in the direction of pump propagation, $L$ is the propagation length, and $e^{-\gamma'(\boldsymbol{k}_s)L}$ is the quantum mechanical signal photon decay rate. In our analysis we consider an effective medium model and take into account dispersion and losses perturbatively; the validity of this approach is discussed in Appendix B, Appendix E, and Appendix F.

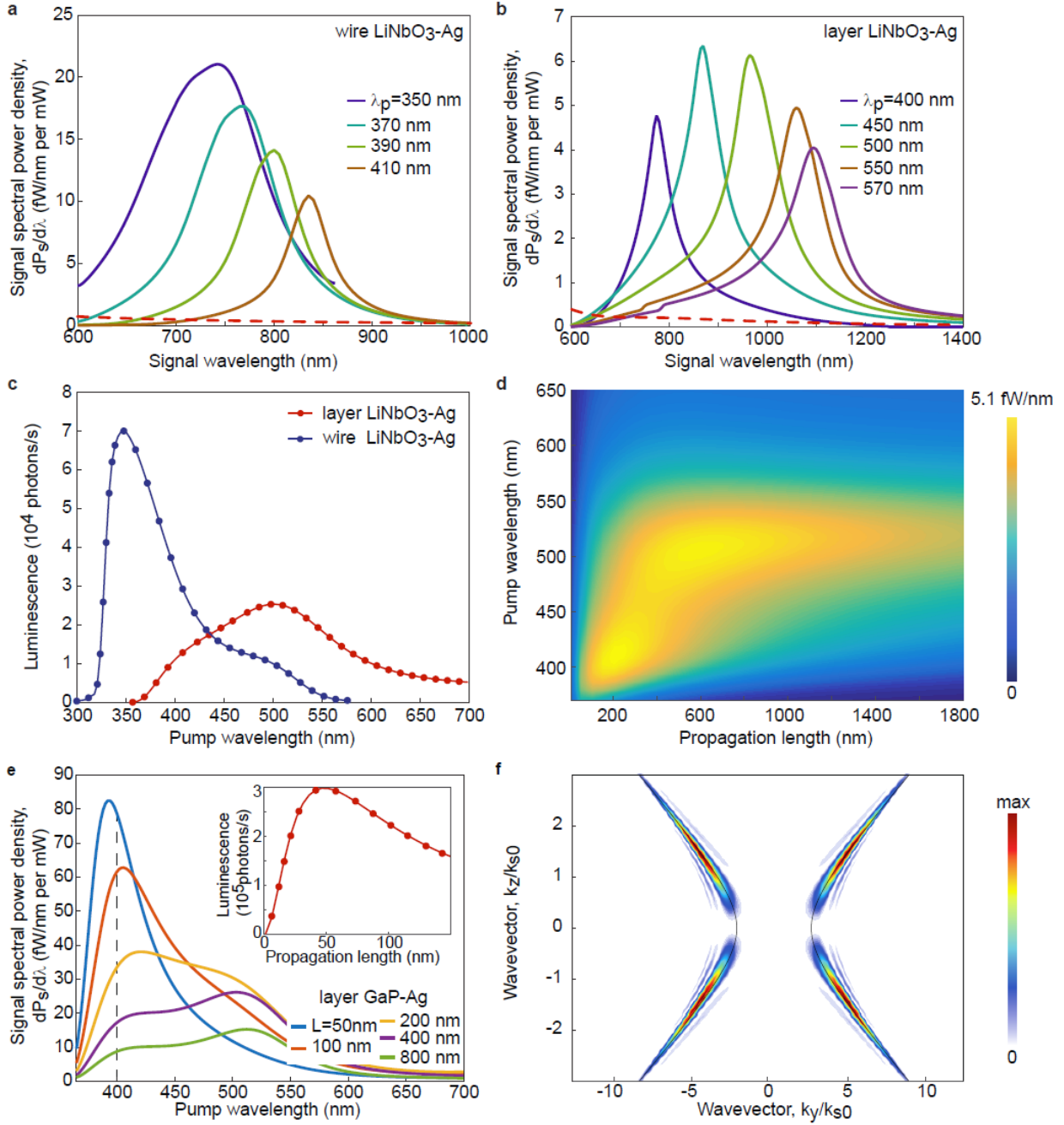

FIG. 4. Broadband parametric downconversion in metamaterials. (a) Calculated spectral power density of emitted signal photons for a wire-like hyperbolic $LiNbO_3 - Ag$ metamaterial for different pump wavelengths. The dashed curve shows expected emission from a bulk $LiNbO_3$ crystal of similar thickness at $\lambda_p = 350$ nm. (b) Spectral power density of emitted signal photons for a layer-hyperbolic metamaterial. The dashed curve shows the expected emission from a bulk $LiNbO_3$ at $\lambda_p = 500$ nm, for comparison. (c) Single photon luminescence rate for frequency-degenerate downconversion (i.e., $2\lambda_p = \lambda_s = \lambda_i$) as a function of a pump wavelength for layered hyperbolic and wire-like hyperbolic $LiNbO_3$–$Ag$ metamaterials. (d) Signal photon spectral power density at the frequency-degenerate downconversion wavelength for a layered hyperbolic metamaterial as function of pump wavelength and propagation length. (e) Calculated signal

*photon spectral power density at the frequency-degenerate downconversion wavelength as a function of the pump wavelength for different propagation lengths for a layer-hyperbolic 70 nm GaP – 30 nm Ag metamaterial. Inset shows the calculated luminescence for a 400 nm pump wavelength. (f) Signal photon emission map for a layer-hyperbolic GaP-Ag crystal at 400 nm pump wavelength after 500 nm of propagation.*

We consider further, as an example, nonlinear parametric downconversion in metal– dielectric hyperbolic metamaterials whose dielectric components comprise a second order nonlinear medium. Figures 4 (a) and 4(b) show the power emission spectra calculated for silver (Ag) – lithium niobate (LiNbO$_3$) wire-like hyperbolic [Fig. 4(a)] and layered hyperbolic [Fig. 4(b)] metamaterials with 80 nm period and 25% metal filling fraction after $L$=500 nm of propagation for 1 mW of input pump power. Note that in our analysis, we consider real material parameters (see Appendix A and Appendix E). For both of the hyperbolic systems we observe strong nonlinear luminescence in a broad range of pump wavelengths ($\simeq$150 nm of operation bandwidth). The peak emitted signal spectral power density reaches 6 fW/nm for a layered hyperbolic metamaterial and 22 fW/nm for a wire-like hyperbolic metamaterial (see also Appendix D).

We estimate the rate of single photon emission, $R_{meta}$, in a degenerate case (i.e., when $\lambda_s = 2\lambda_p$), see [Fig. 4(c)] and Appendix A. Specifically, we predict 25000 photons/s for layered metamaterials and 70000 photons/s for wire-like metamaterials, respectively. To quantify the emission enhancement we introduce a Purcell-like coefficient for parametric luminescence, $F = R_{meta}(L)/R_{regular}(L)$, where we compare the emission rate of our metamaterials with that of a regular unpoled bulk crystal after the same propagation length. For both wire-like and layered hyperbolic metamaterials, we obtain a nearly $F = 50$ times increase in luminescence intensity.

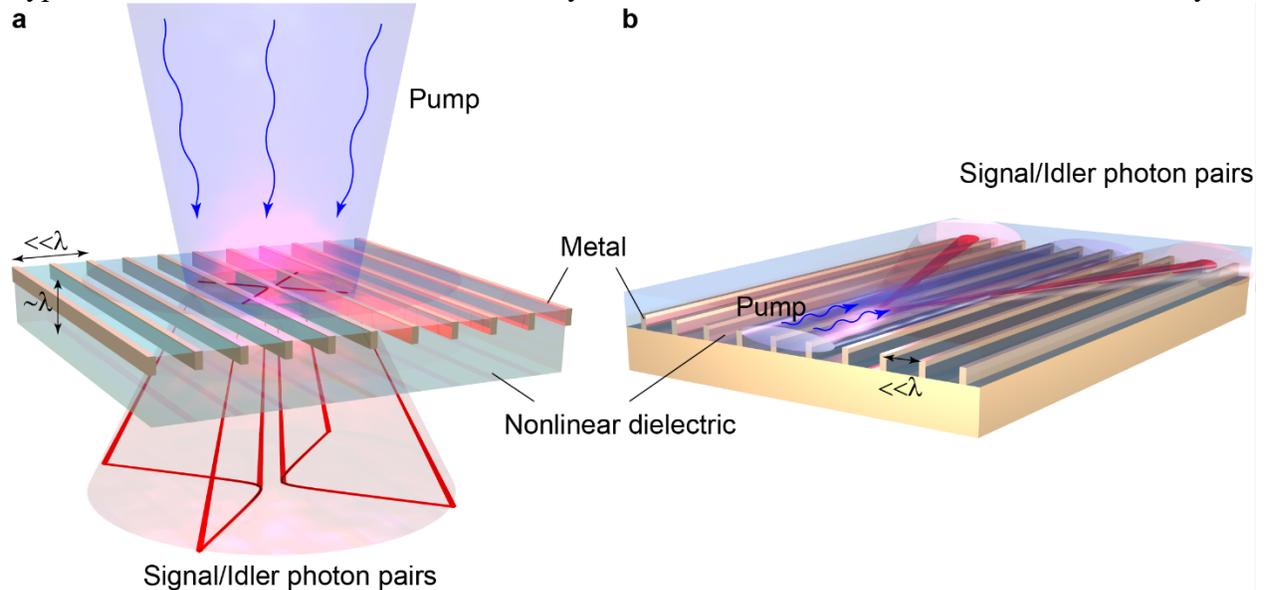

*FIG. 5. Schematic illustration of configurations for possible experimental demonstration of the predicted mismatch-free Purcell enhancement of the nonlinear luminescence in hyperbolic metastructures. As discussed in the text, efficient luminescence is possible after ~100 nm of propagation, suggesting that a metallic grating carved into a nonlinear dielectric substrate with subwavelength spacing and straightforward-to-fabricate aspect ratio (~1/5) would serve as an easy to test system (a). Another proposal is based on a recently demonstrated hyperbolic*

*metasurface [33]; loading it with a nonlinear medium may lead to an on-chip bi-photon generation (b).*

The influence of losses is studied in Fig. 4(d). A peak is clearly seen in emission at $L \simeq 500$ nm due to an interplay between the downconverted photon generation probability, which is proportional to interaction length $L$, and photon absorption, which varies as $e^{-\gamma'(\mathbf{k}_s)L}$). However even after 2-3 $\mu m$ of propagation, the overall luminescence intensity remains reasonably high.

Our predictions may be extrapolated to other second-order materials with refractive indices and nonlinear responses similar to LiNbO$_3$ (the dominant component in LiNbO$_3$ is $d_{33} = 34.4$ pm/V ($\chi^{(2)}_{zzz} = 2d_{33}$) [30]). Hence a nearly 3.5 times weaker signal is expected for a potassium titanyl phosphate (KTP) with $d_{33} = 18.5$ pm/V [31], whereas for some organic polymers $d_{33}$ values as high as $\simeq 100$ pm/V were reported [32]; in this case almost 9 fold stronger effect is anticipated.

Finally, as an example of a different material system, we consider light generation in silver – gallium phosphide (GaP) metamaterials. Gallium phosphide is a high refractive index semiconductor compatible with silicon nanofabrication processes, and demonstrating a strong second order optical response ($d_{24} \simeq 100$ pm/V ($\chi^{(2)}_{xyz} = 2d_{24}$)) [30]. However a significant material dispersion prevents the use of GaP in conventional practice, since the phase matching conditions cannot be met in the bulk crystal. The high index and strong nonlinearity of GaP facilitate compact device integration and higher photon generation rates. Our calculations predict a substantial luminescence at $\lambda_p = 400$ nm after only 50 nm of propagation (about 4 times that of a tenfold thicker LiNbO$_3$-based structures studied above, and $\simeq 1000$ times stronger than for a homogeneous gallium phosphide film of the same thickness), see Fig. 4(e). We anticipate a generation rate of over $3 \times 10^5$ photons/s in this case (see inset in Fig. 4(e)). Light emission pattern shown in Fig. 4(f) is hyperbolic, as is expected from our simple phase matching considerations.

Our predictions of high luminescence intensities in submicron thickness structures suggest that compact broadband nonlinear single photon sources may be designed [see Fig. 5(a)]. We expect that the actual emission enhancement may differ slightly from our predictions due to potential fabrication imperfections and effects beyond the scope of our model (such as a finite pump beam size [17], limitations of the effective medium approach, and boundary effects of the quantization model).

To conclude, in our analysis we have developed a general framework to describe spontaneous nonlinear downconversion in complex three-dimensional metamaterials taking into account photonic band structure, dispersion and losses. We further predicted that in hyperbolic metastructures broadband, enhanced and phase-mismatch-free generation of quantum light may be attained. We note that our theoretical formalism and conceptual approach could be easily extended to other photonic platforms (e.g., hyperbolic metasurfaces [33], Fig. 5(b), and epsilon-near-zero metamaterials [19, 24, 25]), and other frequency domains (near and mid-infrared, where potentially highly nonlinear multiquantum well semiconductor heterostructures may be utilized [18, 34]), as well as nonmetallic systems (such as, all-dielectric near-zero-index crystals [35] and phonon-polariton systems [29]).

## ACKNOWLEDGMENTS

This work was supported by the Air Force Office of Scientific Research through award FA9550-16-1-0019. Authors thank G. Papadakis, R. Sokhoyan, Yu. Tokpanov, M. Alam and O. Ilic for useful discussions

## APPENDIX A: METHODS

Material parameters for silver are taken from experimental data [36].

In our analysis we have assumed that the anisotropy of LiNbO$_3$ is much weaker than that of a hyperbolic metamaterial. With this assumption we modeled the permittivity of LiNbO$_3$ as effective isotropic function being an average between its ordinary and extraordinary permittivity components. The crystal axis of LiNbO$_3$ is considered to be co-aligned with that of hyperbolic metamaterial.

For GaP we assumed a 45 degree rotation of GaP crystal axis with respect to that of a hyperbolic metamaterial – this is a standard consideration that ensures maximum of nonlinear interaction and wavemixing.

The single photon emission rate was estimated assuming a 1 nm spectral filter at the degenerate frequency, i.e., $R = \frac{1}{\hbar\omega}\int\left(\frac{dP}{d\lambda}\right)d\lambda$.

## APPENDIX B: HYPERBOLIC METAMATERIAS AND EFFECTIVE MEDIUM THEORY. BACKGROUND

Progress in nanotechnology, materials science and nanofabrication over recent years has led to design of photonic structures with feature sizes much smaller than the wavelength of the incident light. The electromagnetic properties of such structures may be described by an effective medium approximation, where the optical response is averaged over the light wavelength [37]. It is therefore the cumulative composition of the entire structure, rather than individual structural elements, that determines lightwave propagation in the medium [38]. Importantly, one may design the electromagnetic properties of such effective materials (metamaterials) by controlling the period, element size, and composition of the different materials (e.g., metals with dielectrics) in the structure. With such a design, an electromagnetic response that is typically not attainable in natural materials can be obtained and used for controlling light-matter interaction [37,38].

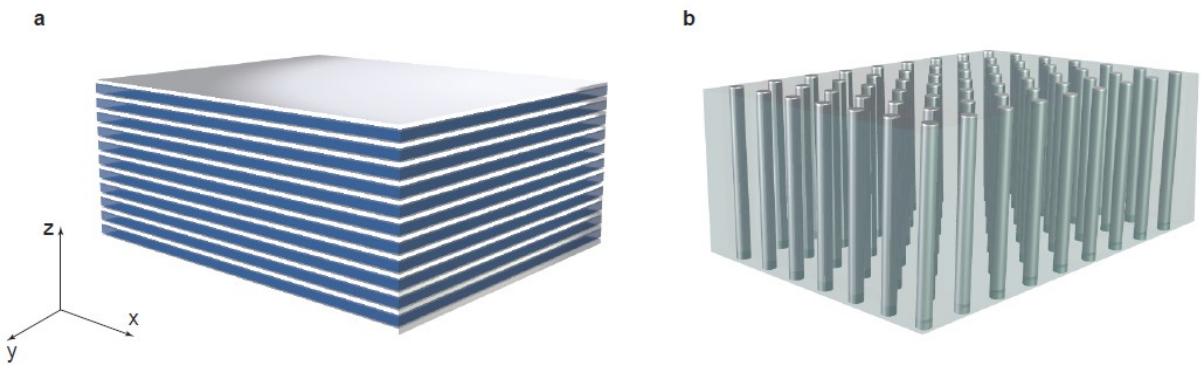

FIG. 6. Schematic illustration of hyperbolic metamaterials. (a) layered hyperbolic media and (b) wire hyperbolic media

An exciting class of structures is that of so-called hyperbolic metamaterials [27, 28, 39]. Hyperbolic metamaterials are periodic metal-dielectric structures, as schematically show in Figs. 6(a) and 6(b) (one may find several comprehensive reviews on the subject, see e.g., [27, 28, 39]). As discussed in the main text, these materials behave on average as uniaxial crystals [40] with negative components in their permittivity tensors. Within the frame of the effective medium theory (i.e., when the structure period is much smaller that the wavelength, $\Lambda \ll \lambda$, [37, 38]) the response may be described as [27]:

$$\varepsilon_o = \varepsilon_\perp = \rho \varepsilon_m + (1-\rho)\varepsilon_d \tag{B1}$$

$$\varepsilon_e = \varepsilon_\parallel = \frac{\varepsilon_m \varepsilon_d}{\rho \varepsilon_d + (1-\rho)\varepsilon_m}$$

in the case of a layered hyperbolic medium [Fig.6 (a)], and as

$$\varepsilon_o = \varepsilon_\perp = \frac{(1+\rho)\varepsilon_m \varepsilon_d + (1-\rho)\varepsilon_d^2}{(1+\rho)\varepsilon_d + (1-\rho)\varepsilon_m} \tag{B2}$$

$$\varepsilon_e = \varepsilon_\parallel = \rho \varepsilon_m + (1-\rho)\varepsilon_d$$

in the case of a wire-like hyperbolic medium [Fig. 6(b)]. In these expressions $\rho$ denotes metal filling fraction (for instance, for a layer medium $\rho = \frac{h_d}{h_d + h_m}$, where $h_d$ and $h_m$ are dielectric and metal layer thicknesses, respectively).

Light dispersion in hyperbolic structures is then described similarly to regular uniaxial crystals:

$$\frac{k_x^2 + k_y^2 + k_z^2}{\varepsilon_\perp} = \frac{\omega^2}{c^2} \quad \text{for ordinary waves} \tag{B3}$$

$$\frac{k_x^2 + k_y^2}{\varepsilon_\parallel} + \frac{k_z^2}{\varepsilon_\perp} = \frac{\omega^2}{c^2} \quad \text{for extraordinary waves}$$

As was mentioned in the main text, depending on signs of the ordinary and extraordinary permittivities different dispersion regimes may be obtained; see Fig. 2 (a) in the main text.

### APPENDIX C: CLASSICAL PHASE MATCHING ANALYSIS

In a classical nonlinear system, the efficiency of nonlinear interaction and wavemixing critically depends on phase matching between the propagating waves. In this section, we derive and analyze phase matching conditions for a nonlinear hyperbolic metamaterial based on the dispersion relations for ordinary and extraordinary waves [Eqs. (B3)]. For simplicity, we consider degenerate wavemixing, i.e., $\omega_s = \omega_i = \frac{\omega_p}{2}$. We also assume two specific scenarios of pump propagation: along the crystal axis (i.e., $\boldsymbol{k}_p \parallel \boldsymbol{z}$) and perpendicular to it (i.e., $\boldsymbol{k}_p \parallel \boldsymbol{x}$). Finally, we limit our analysis to the case of downconversion into extraordinary signal and extraordinary idler waves (other cases, such as conversion into ordinary-ordinary and extraordinary-ordinary waves, are quite similar to those of crystals with regular material dispersion).

***Pump propagation along the crystal axis $\boldsymbol{k}_p \parallel \boldsymbol{z}$***. Assuming that $n_p$ is the effective index of the pump wave in the direction of propagation, we can express the pump wavevector as $k_{pz} = \frac{\omega_p}{c} n_p$. It is convenient to represent the signal and idler wavevectors as $\boldsymbol{k} = \boldsymbol{k}_\perp + \boldsymbol{k}_\parallel$, where $\boldsymbol{k}_\parallel$ is the component of the wavevector parallel to the direction of pump propagation (in this particular case it is the $z$ component) and $\boldsymbol{k}_\perp$ is the component of the wavector that is perpendicular to the pump (the notation not to be mixed with $\varepsilon_\parallel$ and $\varepsilon_\perp$ which are defined with respect to the crystal symmetry axis). Phase matching in a degenerate case, $\omega_s = \frac{\omega_p}{2}$, would require that $2k_{sz} = k_{pz}$, $k_{sz} = k_{iz}$ and $\boldsymbol{k}_{s\perp} = -\boldsymbol{k}_{i\perp}$. Substituting these conditions into the dispersion equations for signal and idler waves, we get:

$$\frac{k_{s\perp}^2}{\varepsilon_{s\parallel}} + \frac{k_p^2}{4\varepsilon_{s\perp}} = \frac{\omega_p^2}{4c^2} = \frac{k_p^2}{4n_p^2} \tag{C1}$$

After a little bit of algebra it is possible to show that:

$$k_{s\perp}^2 = \frac{\omega_s^2}{c^2}\left(\varepsilon_\parallel - n_p^2 \frac{\varepsilon_\parallel}{\varepsilon_\perp}\right) \tag{C2}$$

Exact phase matching in a lossless structure is possible when the transverse signal (idler) wavectors are real, implying that $k_{s\perp}^2 > 0$. For $\varepsilon_\parallel > 0$ and $\varepsilon_\perp > 0$, i.e., in regular crystals, phase matching is possible only when $n_p^2 < \varepsilon_\perp$. This condition is hard to achieve in crystals with normal material dispersion, in which the refractive index monotonically grows with frequency. This effect is especially pronounced in high refractive index structures and semiconductors near the band-gap edge (e.g., gallium phosphide discussed in the main text), challenging their use in nonlinear applications.

In the regime $\varepsilon_\perp < 0$ and $\varepsilon_\parallel > 0$, which is typically attained in a layered hyperbolic media [Fig. 6(a)] phase matching is automatically satisfied. In particular, for any $n_p^2 > 0$ (i.e., for any propagating pump in the crystal) there always exists a pair of signal–idler waves.

Finally, in a wire hyperbolic medium ($\varepsilon_\perp > 0$ and $\varepsilon_\parallel < 0$) phase matching is possible for $n_p^2 > \varepsilon_\perp$, see also Fig. 3(b) in the main text. We note that for $n_p^2 < \varepsilon_\perp$ exact phase matching is satisfied for ordinary waves.

The locus of points in the $\boldsymbol{k}$ space for which phase matching is achieved determines directions of nonlinear light emission, i.e., the angular distribution of the emission pattern. Clearly, for a pump propagating along the $z$ axis, which is the high symmetry axis of the crystal, the emission patterns for all the studied cases are radially symmetric, i.e., $\boldsymbol{k}_{s\perp} = const$, see Figs. 3(a) and 3(b) in the main text.

***Pump propagation along the x axis, $\boldsymbol{k}_p \parallel x$.*** For this direction of pump propagation the phase matching condition requires that $2k_{sx} = k_{px}$. Substituting this condition into the dispersion equation for the extraordinary waves [Eq. (B3)] we arrive at the following expression:

$$k_{sy}^2 + k_{sz}^2 \frac{\varepsilon_\parallel}{\varepsilon_\perp} = \frac{\omega_s^2}{c^2}\left(\varepsilon_\parallel - n_p^2\right) \tag{C3}$$

Phase matching is possible when propagating signal (idler) solutions exist, i.e., $k_{sy}^2 > 0$ and $k_{sz}^2 > 0$. For a regular crystal ($\varepsilon_\parallel \varepsilon_\perp > 0$), this condition is satisfied only for $n_p^2 < \varepsilon_\parallel$, which is again hard to obtain in materials with normal dispersion. The emission pattern in this case is elliptical, $k_{sy}^2 + k_{sz}^2 \frac{\varepsilon_\parallel}{\varepsilon_\perp} = const$.

In the hyperbolic regime, when $\varepsilon_{s\perp}\varepsilon_{s\parallel} < 0$, we find that the phase matching condition is always satisfied. That is, one can always find $k_{sy}$ and $k_{sz}$ that satisfy Eq. (C3). Interestingly, in this scenario light emission is hyperbolic (see. Figs. 3(c) and 3(d) in the main text). Furthermore, for a layered hyperbolic medium ($\varepsilon_{s\perp} < 0$ and $\varepsilon_{s\parallel} > 0$) there is a transition point at $n_p^2 = \varepsilon_\parallel$, at which the emission goes from a single hyperbolic to a double hyperbolic shape (this is easy to see from the geometry of the isofrequency surface, as shown in Fig. 7(f)).

## APPENDIX D: PRINCIPLES OF HYPERBOLIC EMISSION

In the main text we showed that in the hyperbolic regime of operation signal photon emission rate may be substantially enhanced as compared to the case of a bulk crystal. In particular, we have calculated signal photon spectral power density, which we found to be expressed as:

$$\frac{dP_s}{d\lambda_s} = \frac{\hbar \pi c^3 L^2}{\lambda_s^4 \lambda_i} \frac{P_p}{\varepsilon_0 n_p} \int d^2 k_{s\perp} \frac{\partial k_{s\parallel}}{\partial \omega} \frac{\partial k_{i\parallel}}{\partial \omega} c_{k_s}^2 c_{k_i}^2 \times$$
$$N(k_s, k_i) \left|\frac{1-e^{i\Delta k_\parallel L}}{i\Delta k_\parallel L}\right|^2 e^{-i\gamma'(k_s)L} \tag{D1}$$

where $N(k_s, k_i) = \left|\sum_{lmn} \bar{\bar{\chi}}^{(2)}_{lmn} u_l(\omega_p) u_m(\omega_s) u_n(\omega_i)\right|^2$.

This expression is rather complex to analyze directly. On the other hand, in the theory of nonlinear optics of crystals it is frequently assumed that the nonlinear wave interaction and mixing may be described by some effective nonlinearity, which mathematically implies that $N(k_s, k_i) \to const$. In this limit the integral simplifies to:

$$G(\omega_s, \omega_p) = \int d^2 k_{s\perp} c_{k_s}^2 c_{k_i}^2 \frac{\partial k_{s\parallel}}{\partial \omega} \frac{\partial k_{i\parallel}}{\partial \omega} \left|\frac{1-e^{i\Delta k_\parallel L}}{i\Delta k_\parallel L}\right|^2 e^{-i\gamma'(k_s)L} \tag{D2}$$

and depends on the signal and idler wave dispersions (embedded into coefficients $c_{k_s}^2 c_{k_i}^2$), the group velocities in the direction of pump propagation ($\frac{\partial k_{s\parallel}}{\partial \omega}$ and $\frac{\partial k_{i\parallel}}{\partial \omega}$), phase matching ($\left|\frac{1-e^{i\Delta k_\parallel L}}{i\Delta k_\parallel L}\right|^2$), and the signal photon dissipation ($e^{-i\gamma'(k_s)L}$).

In this section, we explore the dynamics of this integral in the hyperbolic regime. For the sake of simplicity of our analysis, and without loss of generality, we consider hyperbolic structures comprised of a nondispersive dielectric with a permittivity $\varepsilon_d = 2$ and a Drude metal with $\varepsilon_m = 1 - \frac{\omega_{plasm}^2}{\omega(\omega - ig)}$, where $\omega_{plasm}$ is the plasma frequency and $g$ is the collision frequency; the metal filling fraction is assumed to be 30% (i.e., $\rho = 0.3$). It is convenient to introduce a wavelength $\lambda_{ENZ}$ for which $\rho\varepsilon_m + (1-\rho)\varepsilon_d = 0$, i.e., at which either ordinary component (for layer-hyperbolic) or extraordinary component (for wire-hyperbolic) of the permittivity tensor go through the epsilon-near-zero point (ENZ) (see Eqs. (B1) and (B2)).

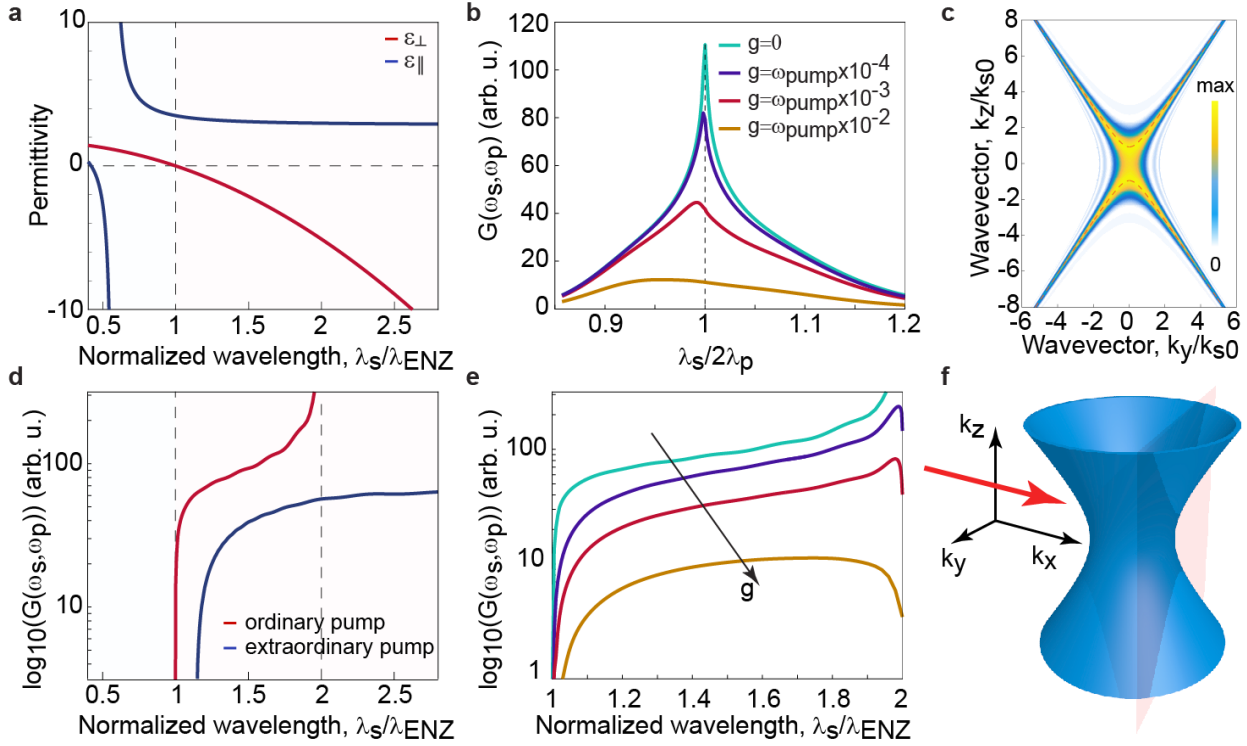

FIG. 7. Spontaneous parametric downconversion in Drude layered hyperbolic medium. (a) dispersion of the ordinary and extraordinary permittivities of a Drude layered hyperbolic medium. (b) spectral variation of the function $G(\omega_s, \omega_p)$ for several different values of the collision frequency, $g$. Here $\lambda_p = \lambda_{ENZ}/1.2$. (c) k-space distribution of the expression under the integral (D2) showing the angular variation of the single photon probability. Here it is assumed that $g = 10^{-3}\omega_{plasma}$ and $\lambda_p = \lambda_{ENZ}/0.9$. Dashed lines denote the emission patterns estimated from the simple phase mateching analysis. (d) logarithmic scale variation of the function $G(\omega_s, \omega_p)$ in a degenerate regime ($\lambda_s = 2\lambda_p$) for two different pump polarizations in a lossless case ($g = 0$). (e) logarithmic scale variation of the function $G(\omega_s, \omega_p)$ in a degenerate regime for ordinary pump polarization for different values of the collision frequency, $g$. In panels (b-e) propagation length is assumed to be $L = 3\lambda_{ENZ}$. (f) schematic illustration of the extraordinary wave isofrequency contour for a layered hyperbolic medium. The arrow denotes the direction of the pump propagation. The cross-section plane schematically explains the origin of the emission pattern observed in panel (c).

Figures 7(a) and 8(a) show dispersion plots for effective ordinary and extraordinary components of the permittivity tensor for layer and wire hyperbolic media, respectively. For wavelengths longer than $\lambda_{ENZ}$ in both of these cases, the system is in the hyperbolic dispersion regime (i.e., $\varepsilon_\parallel \varepsilon_\perp < 0$). As was discussed in the main text, we consider pumping in the elliptical and emission in the hyperbolic parts of the spectrum (Fig. 2(b) in the main text). A pump wave propagating in the $x$ direction (perpendicular to the crystal axis) can have two possible polarization states: ordinary ($k_{pz} = \frac{\omega}{c}\sqrt{\varepsilon_\perp}$, $n_p = \sqrt{\varepsilon_\perp}$) and extraordinary ($k_{pz} = \frac{\omega}{c}\sqrt{\varepsilon_\parallel}$, $n_p = \sqrt{\varepsilon_\parallel}$). The choice of pump polarization would influence the phase matching conditions.

Typical spectra of the function $G(\omega_s, \omega_p)$ for a layered hyperbolic medium are shown in Fig. 7(b) (here an ordinary pump wave is assumed, $L = 3\lambda_{ENZ}$, and $\lambda_p = \lambda_{ENZ}/1.2$). For a lossless case, the emission is peak exactly at the degenerate wavelength $\lambda_s = \lambda_p/2$. With increase losses in the system (i.e., increase of $g$), the emission peak shifts to shorter wavelengths. We attribute this break of symmetry to the frequency dispersion of losses in the system (losses are more pronounced closer to the ENZ point). A similar trend is also seen in the Fig. 4(b) in the main text.

In Fig. 7(d) we plot the function $G(\omega_s, \omega_p)$ at the degenerate wavelength ($\lambda_s = 2\lambda_p$) in a lossless limit. At the edge of the elliptical-to-hyperbolic transition function $G(\omega_s, \omega_p)$ vanishes (we note that at this wavelength rage, close to the near-zero-epsilon point, our perturbative approach might not be fully sufficient). With increasing pump wavelength, i.e., with getting deeper into the hyperbolic domain for signal waves, the function $G(\omega_s, \omega_p)$ dramatically increases, manifesting a dramatic increase in the light emission in the hyperbolic regime. Such an enhancement is associated with phase mismatch free operation and a dramatic growth of the available signal wave phase volume (we again assume here that the maximum wavevector is bound, $k_{zmax} = 10\frac{2\pi}{\lambda_{ENZ}}$, so that the phase volume is always finite). Finally, for the ordinary pump, as $n_p = \sqrt{\varepsilon_\perp} \to 0$, $G(\omega_s, \omega_p)$ diverges. This divergence is essentially a mathematical artifact attributed to the lossless ENZ medium (to sustain a finite pump power in the system an infinitely high electric field is required; this is unphysical). However, in a realistic system the growth of $G(\omega_s, \omega_p)$ is limited by losses in the system, as is clearly shown in Fig. 7(e). Importantly, the function $G(\omega_s, \omega_p)$ is enhanced over a broad range of pump (signal) wavelengths in the hyperbolic regime. A similar dynamics is observed in the case of structures with realistic material parameters (see Fig. 4 in the main text). In particular, we see an initial growth of the peak emission right after the ENZ point for signal waves followed by a broad range of enhanced light emission, and, finally, by a subsequent decrease in the light emission close to the ENZ point for the pump wave (see Figs. 4(a) and 4(b) in the main text).

Next we analyze the angular distribution of light emission in such an idealistic Drude layered hyperbolic medium. The corresponding emission pattern is shown in the Fig. 7(c). The shape of the emission pattern is hyperbolic and coincides with our simple phase matching analysis. Fringes in the pattern are characteristic of the $\text{sinc}(\Delta kL)$ at small propagation distances. As $L \to \infty$ the emission pattern collapses to a curve $\Delta \mathbf{k} = 0$. We find that the maximum emission occurs for smaller signal (idler) wavevectors, $\mathbf{k}$, and rapidly decreases with the $\mathbf{k}$ increase. This behavior is associated with a stronger damping for higher $\mathbf{k}$ modes, as expected. Strong emission for $|\mathbf{k}_\perp|/|\mathbf{k}_{s0}| < 3$ suggests that the emitted signal light may be efficiently collected by adjacent high index dielectric media. A similar hyperbolic emission pattern is shown for the case of a silver – gallium phosphide layered hyperbolic medium (Fig. 4(f) in the main text). However in that case, the influence of the $N(\mathbf{k}_s, \mathbf{k}_i)$ term under the integral (i.e., mixing of light polarizations) modifies the emission pattern.

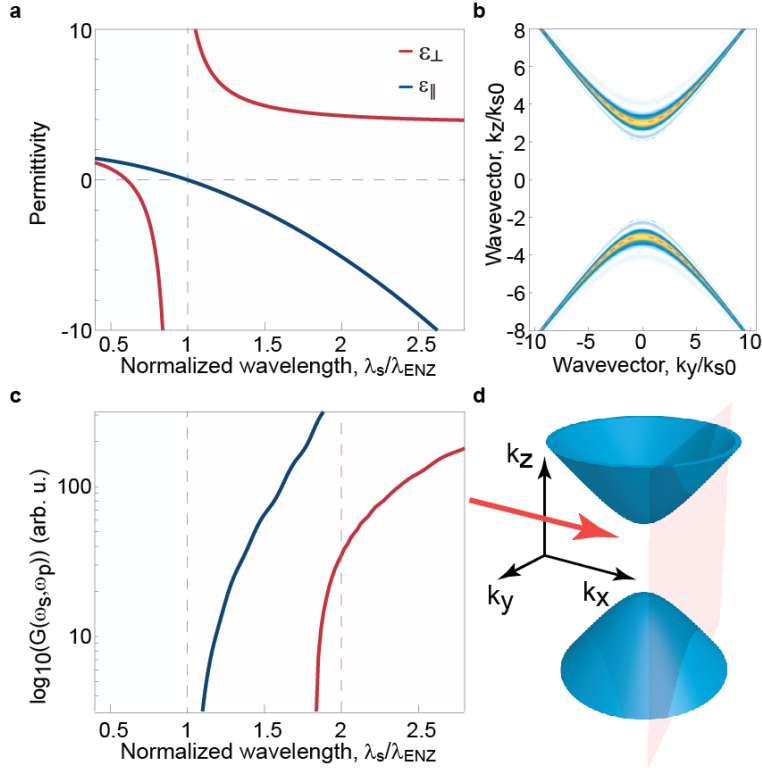

FIG. 8. Spontaneous parametric downconversion in a wire hyperbolic medium. (a) dispersion of the ordinary and extraordinary permittivities of a Drude layered hyperbolic medium. (b) k-space distribution of the expression under the integral (D2) showing the angular variation of the single photon probability. Here it is assumed that $g = 10^{-3}\omega_{plasma}$ and $\lambda_p = \lambda_{ENZ}/0.9$. Dashed lines denote the emission patterns estimated from the simple phase matehing analysis. (c) logarithmic scale variation of the function $G(\omega_s, \omega_p)$ in a degenerate regime ($\lambda_s = 2\lambda_p$) for two different pump polarizations in a lossless case ($g = 0$). (d) schematic illustration of the extraordinary wave isofrequency contour for a wire hyperbolic medium. The arrow denotes the direction of the pump propagation. The cross-section plane schematically explains the origin of the emission pattern observed in panel (b).

The case of a hyperbolic wire-like medium is studied in Fig. 8. Similar to the layer-medium case, the emission (function $G(\omega_s, \omega_p)$) grows dramatically with increasing pump wavelength (i.e., as signal wavelength is pushed deeper into the hyperbolic wavelength range), Fig. 8(c). The emission pattern is double hyperbolic [Fig. 8(b)], as expected from the phase matching analysis, see Figs. 3(c) and 3(d) in the main text. Similarly to the layer-hyperbolic medium, higher $k$-wavevectors are strongly attenuated.

## APPENDIX E: HYPERBOLC Ag-LiNbO₃ AND Ag-GaP STRUCTURES

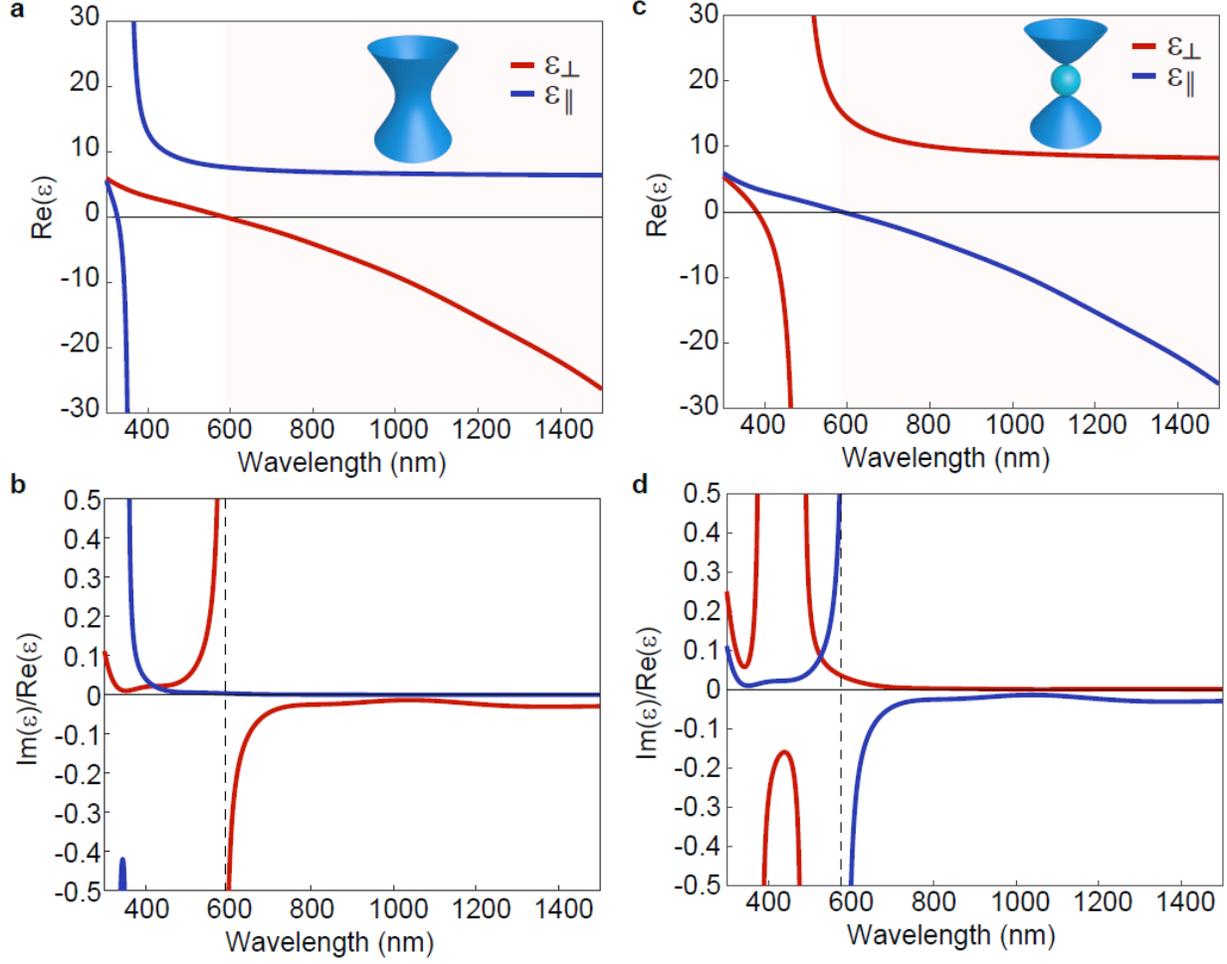

FIG. 9. Effective medium parameters for LiNbO₃ – Ag hyperbolic media with 80 nm period and 25% metal filling fraction. (a) dispersion of the real part of the ordinary and extraordinary components of the effective permittivity tensor for a layer hyperbolic medium. Inset shows the isofrequency surface. (b) Dispersion of the material "quality factor" defined as the ratio of the imaginary to real part of the permittivity. (c) and (d), same as (a) and (b) but for a wire hyperbolic metamaterial.

We next analyze the effective material parameter dispersion for the structures studied in the main text. In Figs. 9(a) and (b), we plot the dispersion of ordinary ($\varepsilon_\perp$) and extraordinary ($\varepsilon_\parallel$) components of the effective permittivity tensor [Eq. (B1) and Eq. (B2)] for layer-hyperbolic [Fig. 9(a)] and wire-hyperbolic [Fig. 9(b)] silver – lithium niobate structures. We consider here a structure with 80nm period and a 25% metal filling fraction; we also take into account actual experimental and material parameters (for LiNbO₃ we used Ref. [41] and for GaP Ref. [42]). For both of the structures depicted, there is an epsilon-near-zero point around λ=600 nm, i.e., for longer wavelengths the system is in the hyperbolic dispersion band, whereas for shorter wavelengths it exhibits elliptical dispersion. The fact that $\lambda_{ENZ}$ is much larger than the period structure (which here is 80 nm) justifies the use of the effective medium theory. Note that the epsilon-near-zero point may be tuned by appropriate choice of the structure period and filling fraction.

In order to estimate the influence of losses in the structure, we plot $\text{Im}(\varepsilon_\perp)/\text{Re}(\varepsilon_\perp)$ and $\text{Im}(\varepsilon_\parallel)/\text{Re}(\varepsilon_\parallel)$ for both layer-hyperbolic [Fig. 9(c)] and wire-like hyperbolic [Figs. 9(d)] structures. These functions physically correspond to the 'quality factor' of the medium. We find that away from the epsilon-near-zero points $\left|\frac{\text{Im}(\varepsilon)}{\text{Re}(\varepsilon)}\right| < 0.1$ implying that within the hyperbolic regime of interest the wave attenuation is not as strong. For instance, for the $x$ direction of propagation for ordinary waves we get $k_x = k'_x + ik''_x = \frac{\omega}{c}\sqrt{\varepsilon_\perp} \simeq \frac{\omega}{c}\sqrt{\text{Re}(\varepsilon_\perp)} + i\frac{\omega}{2c}\sqrt{\text{Re}(\varepsilon_\perp)}\frac{\text{Im}(\varepsilon_\perp)}{\text{Re}(\varepsilon_\perp)}$. The propagation length in terms of light wavelength in the medium is then simply $\mathcal{L} = \frac{k'_x}{2k''_x} = \frac{\text{Re}(\varepsilon_\perp)}{\text{Im}(\varepsilon_\perp)}$, which is according to Figs. 9(c) and 9(d) is over 30 wavelengths for a layer-hyperbolic and over 100 wavelengths for a wire-hyperbolic media. Similar estimates may be carried out for extraordinary waves.

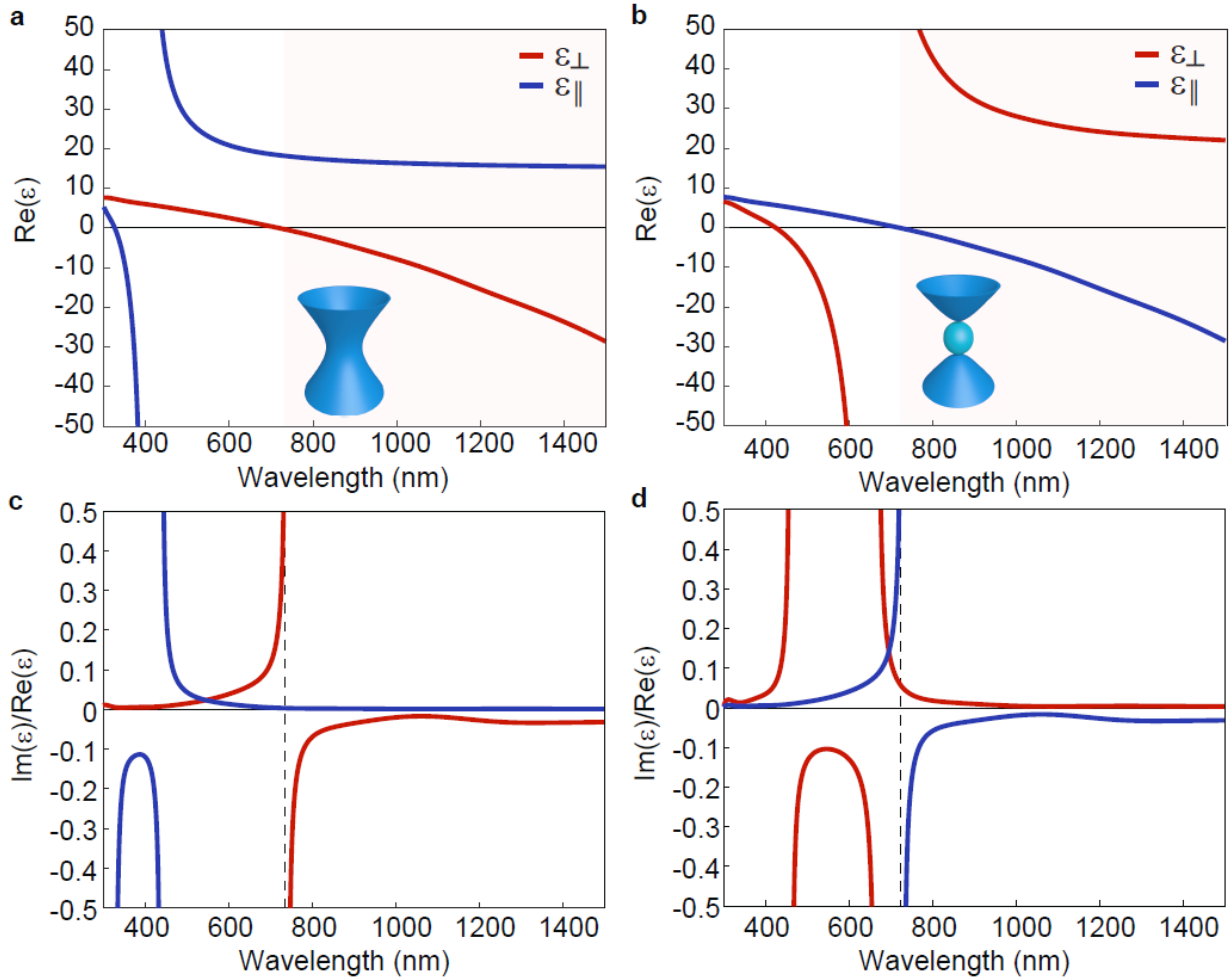

*FIG. 10. Effective medium parameters of GaP – Ag hyperbolic media with a 100 nm period and 30% metal fill fraction. (a) dispersion of ordinary and extraordinary components of the effective permittivity tensor for a layer hyperbolic medium. Inset shows the isofrequency surface. (b) Dispersion of the material "quality factor" defined as the ratio of the imaginary to real part of the permittivty. (c) and (d) same as (a) and (b) but for a wire hyperbolic metamaterial.*

We apply a similar analysis to study the case of gallium phosphide based hyperbolic media, see Fig. 10. Here we assume that the period of the structure is 100 nm and the metal filling fraction is 30%. The elliptical-to-hyperbolic transition is at around λ=750 nm, see Figs. 10 (a) and 10(b). Finally, since GaP is practically lossless above 450 nm, in the hyperbolic regime we obtain material-dependent quality factors similar to those of Ag-LiNbO$_3$ structures, Figs. 10(c) and 10(d).

### APPENDIX F: SPONTANEOUS PARAMETRIC DOWNCONVERSION EMISSION RATE AND POWER.

In this section, we develop a quantum theory that allows us to estimate the probability of signal photon generation through spontaneous parametric downconversion in uniaxial crystals with quadratic nonlinearity, and dispersive and lossy effective permittivity tensor $\bar{\bar{\varepsilon}} = \mathrm{diag}(\varepsilon_\perp, \varepsilon_\perp, \varepsilon_\parallel)$. We develop a theory that may be capable of describing SPDC in either of the scenarios shown in Fig. 1(c) of the main text: regular elliptical (with loss and dispersion), layer-hyperbolic and wire-hyperbolic cases.

In principle, there are two approaches in developing such a theory: microscopic and macroscopic [43-51]. In a microscopic picture one considers both electromagnetic field and matter degrees of freedom and, thus, accounting explicitly for light-matter interaction, dissipation, and dispersion [46-48]. However, such a formulation in case of an extreme anisotropy and material nonlinearity becomes rather cumbersome, and hard to analyze and implement. The macroscopic picture is a standard formalism for quantum optics of dielectric media [45,50]. Within the scope of this theory a classical macroscopic electromagnetic field is quantized. Previously, macroscopic theory was utilized to describe spontaneous parametric downcovesion in a quasi-isotropic, lossless, and dispersionless limit [51-53]. Ref. [26] proposed a Green's function formalism for accounting spontaneous nonlinear wavemixing in complex systems. Here we develop a conceptually different framework for accounting for spontaneous parametric downconversion in a general case of lossy and dispersive medium with arbitrary anisotropy, which is more intuitive, in many ways straightforward and that may provide deeper insight into the physics of quantum nonlinear processes. Specifically, we base our analysis on a macroscopic picture, considering eigen-mode configuration and take into account material losses and dispersion perturbatively.

We consider that the electric field in the crystal may be presented as a sum of interacting pump, signal and idler waves: $\boldsymbol{E} = \boldsymbol{E}_p + \boldsymbol{E}_s + \boldsymbol{E}_i$. We treat the pump classically and explore the quantum mechanical spontaneous generation of signal and idler photons. For this reason, we write down the Hamiltonian of the system:

$$\widehat{\mathcal{H}} = \widehat{\mathcal{H}}_{lin} + \widehat{\mathcal{H}}_{int} \tag{F1}$$

where $\hat{\mathcal{H}}_{lin}$ is the linear part of the Hamiltonian corresponding to the self-energy of the interacting fields, and $\hat{\mathcal{H}}_{int}$ is the energy of the system associated with the nonlinear interaction between signal, idler and pump waves.

The self-energy associated with either of the interacting fields to the first order approximation of the dispersion theory may be expressed as [17,54]:

$$\hat{\mathcal{H}}_{lin}^{s,i} = \frac{1}{2}\int_V d^3r \left[\frac{1}{\mu_0}\hat{B}^2 + \varepsilon_0 \hat{E}\bar{\bar{\varepsilon}}\hat{E} + \varepsilon_0 \omega \hat{E} \frac{\partial \bar{\bar{\varepsilon}}}{\partial \omega}\hat{E}\right]_{s,i} \tag{F2}$$

here the field operators $\hat{B}$ and $\hat{E}$, and material parameters are to be taken at signal or idler wave frequencies, respectively, the integration is over the quantization volume of the crystal $V = A \times L$, where $A$ is the crystal area perpendicular to the pump propagation direction and $L$ is the distance along the propagation direction (i.e., interaction length).

The interaction Hamiltonian, in turn, is given by the following expression [55-57]:

$$\hat{\mathcal{H}}_{int} = \varepsilon_0 \int_V d^3r \sum_{lmn} \bar{\bar{\chi}}^{(2)}_{lmn} E_l(\omega_p) E_m^+(\omega_s) E_n^+(\omega_i) + H.c. \tag{F3}$$

where $\bar{\bar{\chi}}^{(2)}$ is the second order nonlinear susceptibility tensor of the crystal, which is assumed to be nondispersive and possessing Klienman's symmetry [57].

We use a standard approach to field quantization, in which the interacting signal and idler fields are decomposed into plane eigen-modes of the crystal [43]. In this case the electric field operator is given as.

$$\hat{E}_\xi = \frac{1}{\sqrt{V}}\sqrt{\frac{\hbar \omega_\xi}{\varepsilon_0}} \sum_\nu \sum_{k_{\xi\nu}} c_{k_{\xi\nu}} u_{k_{\xi\nu}} \hat{a}_{k_{\xi\nu}} e^{ik_{\xi\nu}r - i\omega_\xi t} + H.c. \tag{F4}$$

where $\hat{a}_{k_{\xi\nu}}$ and $\hat{a}^+_{k_{\xi\nu}}$ are photon annihilation and creation operators, respectively, index $\xi = \{s,i\}$ differentiates signal and idler waves, index $\nu$ runs over ordinary and extraordinary waves that may exist in the uniaxial crystal. The electric field polarizations $u_{k_{\xi\nu}}$ of the crystal in a general case are given as:

$$u = \begin{bmatrix} u_x \\ u_y \\ u_z \end{bmatrix} = u_0 \begin{bmatrix} \dfrac{k_x}{k^2 - \dfrac{\omega^2}{c^2}\varepsilon_\perp} \\ \dfrac{k_y}{k^2 - \dfrac{\omega^2}{c^2}\varepsilon_\perp} \\ \dfrac{k_z}{k^2 - \dfrac{\omega^2}{c^2}\varepsilon_\parallel} \end{bmatrix} \tag{F5}$$

where we dropped indexes for simplicity. The normalization is chosen such that polarizations satisfy orthonormality condition $u_{k_{\xi\nu}} u'_{k_{\xi\nu}} = \delta_{\nu\nu'}$. A special treatment is needed when $k^2 =$

$\frac{\omega^2}{c^2}\varepsilon_{\perp,\parallel}$. Thus for the case of ordinary waves $k^2 = \frac{\omega^2}{c^2}\varepsilon_\perp$ polarization reduces to $\boldsymbol{u} = \boldsymbol{u}_0(k_y, -k_x, 0)$, whereas for extraordinary waves, when $k^2 = \frac{\omega^2}{c^2}\varepsilon_\parallel$ polarization is given simply as $\boldsymbol{u} = (0,0,1)$. The wavevectors $\boldsymbol{k}_{\xi\nu}$ for a given frequency $\omega_\xi$ may be found from the respective dispersion equations for ordinary and extraordinary waves, see Eq. (B3). Lastly, in Eq. (F4) normalization coefficients $c_{\boldsymbol{k}_{\xi\nu}}$ are chosen such that the energy of each of the eigen-modes $\boldsymbol{k}_{\xi\nu}$ is quantized in the units of $\hbar\omega_\xi$:

$$c_{\boldsymbol{k}_{\xi\nu}} = \left[\boldsymbol{u}_{\boldsymbol{k}_{\xi\nu}}\bar{\bar{\varepsilon}}_\xi \boldsymbol{u}_{\boldsymbol{k}_{\xi\nu}} + \frac{1}{2}\boldsymbol{u}_{\boldsymbol{k}_{\xi\nu}}\omega_\xi \frac{\partial\bar{\bar{\varepsilon}}_\xi}{\partial\omega}\boldsymbol{u}_{\boldsymbol{k}_{\xi\nu}}\right]^{-1/2} \tag{F6}$$

The pump field is considered to be classical and is expressed as:

$$\boldsymbol{E}_p = \frac{1}{2}\sqrt{\frac{2P_p}{\varepsilon_0 c n_p A}}\boldsymbol{u}_p e^{i\boldsymbol{k}_p\boldsymbol{r}-i\omega t} + c.c. \tag{F7}$$

where $n_p$ is the effective index of the pump wave along the direction of pump propagation, $P_p$ is the pump power in the crystal.

Next, we use Fermi's Golden rule to estimate the single photon generation rate. We assume that the losses are weak and do not perturb the nonlinear interaction and spontaneous photon downconversion. Specifically, we consider that the probabilities of photon emission and subsequent photon absorption are independent of each other. In this case the rate of generation of signal photons with a given frequency $\omega_s$ and wavevector $\boldsymbol{k}_s(\omega_s)$ for all possible idler waves can be written as [17,58]:

$$\mathcal{R}_s(\omega_s, \boldsymbol{k}_s) = \frac{2\pi}{\hbar}\int_{\mathcal{V}_i}\frac{V}{8\pi^3\hbar}|<f|\widehat{\mathcal{H}}_{int}|0>|^2 \delta(\omega_p - \omega_s - \omega_s)d^3\boldsymbol{k}_i \tag{F8}$$

here, as was mentioned in the main text, $|0>$ corresponds to a state containing no signal and idler photons, and $<f|$ is the final state with one signal photon with frequency $\omega_s$ and wavevector $\boldsymbol{k}(\omega_s)$, and one idler photon with frequency $\omega_i = \omega_p - \omega_s$ and wavevector $\boldsymbol{k}(\omega_i)$; the integration is carried out over the entire idler wave phase space $\mathcal{V}_i$. Note that we dropped here ordinary or extraordinary wave notation (index $\nu$) for the sake of simplicity.

The transition matrix element with the use of Eq. (F4) is then:

$$<f|\widehat{\mathcal{H}}_{int}|0> = \hbar\sqrt{\omega_s\omega_i}\frac{1}{V}c_{\boldsymbol{k}_s}c_{\boldsymbol{k}_i}\frac{1}{2}\sqrt{\frac{2P_p}{\varepsilon_0 c n_p A}}$$

$$\times \sum_{lmn}\bar{\bar{\chi}}^{(2)}_{lmn}u_l(\omega_p)u_m(\omega_s)u_n(\omega_i)\int_V d^3\boldsymbol{r}\, e^{i\Delta\boldsymbol{k}\boldsymbol{r}} \tag{F9}$$

where $\Delta\boldsymbol{k} = (\boldsymbol{k}_p - \boldsymbol{k}_{s\parallel} - \boldsymbol{k}_{i\parallel}) - (\boldsymbol{k}_{s\perp} + \boldsymbol{k}_{i\perp})$, $\boldsymbol{k}_p$ is the pump wavevector, $\boldsymbol{k}_{\xi\parallel}$ and $\boldsymbol{k}_{\xi\perp}$ are components of the signal (idler) wavevectors parallel and perpendicular to the pump one, respectively (not to mix with $\varepsilon_\perp$ and $\varepsilon_\parallel$ that are linked with axes of the uniaxial crystal, see Fig.

1(c) of the main text). In the limit of large crystal area, i.e., $A \to \infty$, the expression for the transition matrix element is transformed to:

$$<f|\hat{\mathcal{H}}_{int}|0> = \hbar\sqrt{\omega_s \omega_i}\frac{4\pi^2}{V}c_{k_s}c_{k_i}\frac{1}{2}\sqrt{\frac{2P_p}{\varepsilon_0 c n_p A}} \times$$

$$\sum_{lmn} \bar{\bar{\chi}}^{(2)}_{lmn} u_l(\omega_p) u_m(\omega_s) u_n(\omega_i) \frac{1 - e^{i\Delta k_\parallel L}}{i\Delta k_\parallel} \delta(\mathbf{k}_{s\perp} - \mathbf{k}_{i\perp}) \tag{F10}$$

where $\Delta k_\parallel = (\mathbf{k}_p - \mathbf{k}_{s\parallel} - \mathbf{k}_{i\parallel})\frac{\mathbf{k}_p}{|\mathbf{k}_p|}$.

Let us go back to the rate equation, Eq. (F8). Using a standard Van Hove transformation $d^3\mathbf{k}_i = d^2\mathbf{s}_i \frac{d\omega}{|\nabla_{\mathbf{k}_i}\omega|}$ we arrive at [26]:

$$\mathcal{R}_s(\omega_s, \mathbf{k}_s) = \frac{1}{\hbar^2} \int_{\partial V_i} \frac{V}{4\pi^2} |<f|\hat{\mathcal{H}}_{int}|0>|^2 \frac{1}{|v_g(\mathbf{k}_i)|} d^2\mathbf{s}_i \tag{F11}$$

where the integration is over the idler isofrequency surface $\partial V_i$, for which $\mathbf{k}_i = \mathbf{k}_i(\omega_i = \omega_p - \omega_s)$, $d^2\mathbf{s}_i$ is the isofrequency surface element, and $v_g(\mathbf{k}_i) = \nabla_{\mathbf{k}_i}\omega_i$ is the idler group velocity.

However for our proceeding calculations, in which we consider pump propagation along some given direction, it is convenient to use $d^3\mathbf{k}_i = d^2\mathbf{k}_{i\perp}\frac{\partial k_{i\parallel}}{\partial \omega}d\omega$ instead, where $\mathbf{k}_{i\perp}$ and $\mathbf{k}_{i\parallel}$ are the components of the idler wavevector perpendicular and parallel to the pump propagation direction, respectively. In this case the rate of signal photon emission is found as:

$$\mathcal{R}_s(\omega_s, \mathbf{k}_s) = \frac{1}{\hbar^2}\int V 4\pi^2 |<f|\hat{\mathcal{H}}_{int}|0>|^2 \frac{\partial k_{i\parallel}}{\partial \omega} d^2\mathbf{k}_{i\perp} \tag{F12}$$

here the integration is carried over a an area in the idler wave phase space on a plane $\mathbf{k}_{i\parallel} = 0$ bound by the isofrequency contour $\partial V_i$. $\frac{\partial k_{i\parallel}}{\partial \omega}$ has a meaning of the idler wave group velocity in the direction on the pump propagation.

So far in our quantization of the signal and idler fields, and in the study of their interactions, we have neglected dissipation, having considered real-valued material parameters at signal and idler photon frequencies, i.e., $\bar{\bar{\varepsilon}} \to \text{Re}(\bar{\bar{\varepsilon}})$. Such an approach is justified when $\left|\frac{\text{Im}(\bar{\bar{\varepsilon}})}{\text{Re}(\bar{\bar{\varepsilon}})}\right| \ll 1$, as we have discussed earlier. We note that there is no such restriction on a pump wave, since it is treated classically (that is, we take into full consideration losses at the frequency of the pump). We treat signal and idler photon losses perturbatively. Specifically, we consider the interplay of probabilities of photon generation and subsequent absorption. In a lossless limit the probability of observing a signal photon at time $t$ is $\mathcal{P}'_s = \mathcal{R}_s t$. In the presence of photon dissipation, this expression has to be modified. In particular, the probability of observing a signal photon emitted in the interval of time $\tau + d\tau$ at a later time $t + \tau$ with accounting for the probability of subsequent photon absorption may be found as:

$$\delta \mathcal{P}_s(t,\tau) = \frac{d\mathcal{P}'_s(\tau)}{d\tau} d\tau e^{-\gamma(k_s)(t-\tau)} \tag{F13}$$

where $\gamma(\boldsymbol{k}_s)$ corresponds to the signal photon dissipation rate that will be determined later. Here we assumed also independence of photon generation and absorption events. The overall probability of observing photon at time $t$ is then $\mathcal{P}_s(t) = \int_0^t \delta\mathcal{P}_s(t,\tau)d\tau$. Hence the signal photon generation rate ($\mathcal{R}_s = \frac{d\mathcal{P}_s}{dt}$) with the account of dissipation is modified as:

$$\mathcal{R}_s \to \mathcal{R}_s e^{-\gamma(k_s)t} \tag{F14}$$

here $t$ has a meaning of interaction time, which we estimate as $t = \frac{L}{v_{g\parallel}(k_s)}$, where $L$ is the propagation length and $v_{g\parallel}(\boldsymbol{k}_s)$ is the group velocity of the generated signal photon with vector $\boldsymbol{k}_s$ at the frequency $\omega_s$ in the direction of pump propagation.

Next we estimate the downconverted signal photon power $dP_s$ emitted per frequency interval $d\omega_s$ integrated over all possible emission angles, i.e., we calculate the emitted signal photon spectral power density:

$$\frac{dP_s}{d\omega_s} = \frac{V}{8\pi^3} \int_{\mathcal{V}_s} \hbar\omega_s \mathcal{R}_s(\omega_s, \boldsymbol{k}_s) e^{-\gamma'(\boldsymbol{k}_s)L} \delta(\omega_s) d^3\boldsymbol{k}_s \tag{F15}$$

Here the integration is over entire signal photon phase space $\mathcal{V}_s$ and $\gamma'_s(\boldsymbol{k}_s) = \gamma(\boldsymbol{k}_s)v_{g\parallel}(\boldsymbol{k}_s)$. Using $d^3\boldsymbol{k}_s = d^2\boldsymbol{k}_{s\perp}\frac{\partial k_{s\parallel}}{\partial\omega}d\omega_s$ we find that:

$$\frac{dP_s}{d\omega_s} = \frac{\hbar\omega_s V}{8\pi^3} \int \frac{\partial k_{s\parallel}}{\partial\omega} \mathcal{R}_s(\omega_s, \boldsymbol{k}_s) e^{-\gamma'(\boldsymbol{k}_s)L} d^2\boldsymbol{k}_{s\perp} \tag{F16}$$

Substituting the expressions for $\mathcal{R}(\omega_s, \boldsymbol{k}_s)$ and $<f|\hat{\mathcal{H}}_{int}|0>$ into the Eq. (F16) we arrive to:

$$\frac{dP_s}{d\omega_s} = \frac{\hbar\omega_s^2\omega_i L^2}{2(2\pi)^3} \frac{P_p}{\varepsilon_0 c n_p} \int d^2\boldsymbol{k}_{s\perp} \frac{\partial k_{s\parallel}}{\partial\omega}\frac{\partial k_{i\parallel}}{\partial\omega} c_{ks}^2 c_{ki}^2 \times$$
$$\left|\sum_{lmn} \bar{\bar{\chi}}^{(2)}_{lmn} u_l(\omega_p) u_m(\omega_s) u_n(\omega_i)\right|^2 \left|\frac{1-e^{i\Delta k_\parallel L}}{i\Delta k_\parallel L}\right|^2 e^{-\gamma'(k_s)L} \tag{F17}$$

here it is assumed that transverse phase matching is satisfied (i.e., $\boldsymbol{k}_{i\perp} = -\boldsymbol{k}_{s\perp}$). We note that in the limit of isotropic, dispersionless and lossless system this expression simplifies to that given in Refs. [17, 52].

Finally making a transform $d\omega_s \to d\lambda_s$ we get:

$$\frac{dP_s}{d\lambda_s} = \frac{\hbar\pi c^3 L^2}{\lambda_s^4 \lambda_i} \frac{P_p}{\varepsilon_0 n_p} \int d^2\boldsymbol{k}_{s\perp} \frac{\partial k_{s\parallel}}{\partial\omega}\frac{\partial k_{i\parallel}}{\partial\omega} c_{ks}^2 c_{ki}^2 \times$$
$$\left|\sum_{lmn} \bar{\bar{\chi}}^{(2)}_{lmn} u_l(\omega_p) u_m(\omega_s) u_n(\omega_i)\right|^2 \left|\frac{1-e^{i\Delta k_\parallel L}}{i\Delta k_\parallel L}\right|^2 e^{-\gamma'(k_s)L} \tag{F18}$$

This equation provides important insights into SPDC in a general uniaxial crystal. In particular, the photon emission rate depends on the interplay between the probabilities of SPDC photon generation, which grows linearly with interaction length $L$, and single photon absorption, which exponentially increases with distance, $e^{-\gamma(k_s)L}$. In a nondegenerate case ($\omega_s \neq \omega_i$), emission at the signal wavelength may be tuned by controlling the idler wave dispersion. For instance, stronger signal photon generation is expected in the slow light regime at the idler frequency (when $v_{g\parallel}(k_i) \ll c$).

Next we estimate the decay rate $\gamma(\boldsymbol{k}_s)$ of generated signal photons. The mechanisms of single photon dissipation in metallic and nanophotonic structures are currently an active topic of research (see for instance Ref. [59]). Here we assume the most simple case of a photon coupled with a thermal reservoir (such a picture would correspond physically to an Ohmic-like dissipation in metallic systems within a Drude regime of dispersion). In this case standard quantum Langevin equations for photon annihilation (creation) operators may be derived:

$$\frac{d\hat{a}_{k_s}}{dt} = -i\omega_s \hat{a}_{k_s} - \frac{1}{2}\gamma_{QM}(\boldsymbol{k}_s)\hat{a}_{k_s} + \hat{F}(t)e^{-i\omega_s t} \tag{F19}$$

where $\gamma_{QM}(\boldsymbol{k}_s)$ is the quantum mechanical decay rate that depends on the density of bath states and photon – reservoir coupling, and $\hat{F}(t)$ is the quantum noise operator.

For a reservoir in thermal equilibrium, the photon number expectation value corresponding to the classical electromagnetic field intensity may be found as:

$$\frac{d<\hat{a}_{k_s}^+ \hat{a}_{k_s}>}{dt} = -\gamma_{QM}(\boldsymbol{k}_s)<\hat{a}_{k_s}^+ \hat{a}_{k_s}> + \frac{\gamma_{QM}(\boldsymbol{k}_s)}{2}\frac{1}{e^{\frac{\hbar\omega_s}{kT}}-1} \tag{F20}$$

where the last term plays a role of a diffusion coefficient. At optical frequencies of interest $\hbar\omega_s \gg kT$ and therefore it may be neglected. Hence Eq. (F20) reduces to a well familiar classical equation for the field intensity dissipation, implying that $\gamma_{QM}(\boldsymbol{k}_s) \simeq \gamma(\boldsymbol{k}_s)$. The classical decay rate, $\gamma(\boldsymbol{k}_s)$, in the limit of $\left|\frac{\text{Im}(\varepsilon)}{\text{Re}(\varepsilon)}\right| \ll 1$ may be estimated from a Poynting theorem [22]:

$$\frac{d\hat{\mathcal{H}}_{lin}^s}{dt} = -\varepsilon_0 \omega_s \int_V d^3\boldsymbol{r} \sum_\nu \text{Im}(\varepsilon_\nu(\omega_s))\hat{E}_{s\nu}^2 \tag{F21}$$

here $\nu$ runs over the Cartesian coordinates. After some algebra we get:

$$\gamma(\boldsymbol{k}_s) = \omega_s c_{k_s}^2 \sum_\nu \text{Im}(\varepsilon_\nu(\omega_s))u_\nu^2 \tag{F22}$$

***Integration boundaries.*** Here we give explicit expressions for the integration boundaries in Eq. (F18).

For $\boldsymbol{z}$ directed prorogation of the pump wave, it is convenient to transform to an integral in polar coordinates, i.e., $d^2\boldsymbol{k}_\perp = k_\rho dk_\rho d\phi$. In this case the integration is as follows:

$$\int_0^{\frac{\omega}{c}\sqrt{\varepsilon_\parallel}} k_\rho dk_\rho \int_0^{2\pi} d\phi <> \qquad for \quad \varepsilon_\perp > 0 \quad and \quad \varepsilon_\parallel > 0$$

$$\int_{\frac{\omega}{c}\sqrt{\varepsilon_\parallel}}^{\sqrt{\frac{\omega^2}{c^2}\varepsilon_\parallel - k_{zmax}^2 \frac{\varepsilon_\parallel}{\varepsilon_\perp}}} k_\rho dk_\rho \int_0^{2\pi} d\phi <> \qquad for \quad \varepsilon_\perp < 0 \quad and \quad \varepsilon_\parallel > 0$$

$$\int_0^{\sqrt{\frac{\omega^2}{c^2}\varepsilon_\parallel - k_{zmax}^2 \frac{\varepsilon_\perp}{\varepsilon_\parallel}}} k_\rho dk_\rho \int_0^{2\pi} d\phi <> \qquad for \quad \varepsilon_\perp > 0 \quad and \quad \varepsilon_\parallel < 0$$

For $x$ direction of pump wave propagation we get:

$$\int_{-\frac{\omega}{c}\sqrt{\varepsilon_\parallel}}^{\frac{\omega}{c}\sqrt{\varepsilon_\parallel}} dk_y \int_{-\sqrt{\frac{\omega^2}{c^2}\varepsilon_\perp - k_y^2 \frac{\varepsilon_\perp}{\varepsilon_\parallel}}}^{\sqrt{\frac{\omega^2}{c^2}\varepsilon_\perp - k_y^2 \frac{\varepsilon_\perp}{\varepsilon_\parallel}}} dk_z <> \qquad for \quad \varepsilon_\perp > 0 \quad and \quad \varepsilon_\parallel > 0$$

$$\int_{-k_{zmax}}^{k_{zmax}} dk_z \int_{-\sqrt{\frac{\omega^2}{c^2}\varepsilon_\parallel + k_z^2 \frac{\varepsilon_\parallel}{|\varepsilon_\perp|}}}^{\sqrt{\frac{\omega^2}{c^2}\varepsilon_\parallel + k_z^2 \frac{\varepsilon_\parallel}{|\varepsilon_\perp|}}} dk_y <> \qquad for \quad \varepsilon_\parallel > 0 \quad and \quad \varepsilon_\perp < 0$$

$$2 \int_{\frac{\omega}{c}\sqrt{\varepsilon_\perp}}^{k_{zmax}} dk_z \int_{-\sqrt{-\frac{\omega^2}{c^2}|\varepsilon_\parallel| + k_z^2 \frac{|\varepsilon_\parallel|}{\varepsilon_\perp}}}^{\sqrt{-\frac{\omega^2}{c^2}|\varepsilon_\parallel| + k_z^2 \frac{|\varepsilon_\parallel|}{\varepsilon_\perp}}} dk_y <> \qquad for \quad \varepsilon_\perp > 0 \quad and \quad \varepsilon_\parallel > 0$$

where $k_{zmax} = \frac{2\pi}{\Lambda}$ is the maximum allowed wavelength in the system.

Note that at the integration boundaries $\frac{\partial k_\parallel}{\partial \omega}$ diverge. This issue can be resolved by transforms of the form $k = C\sin(\theta)$ or $k = C\cos(\theta)$, where $C$ is a constant.

***A note on quantization.*** In our analysis, we have used a common approach of quantizing an electric field $\boldsymbol{E}$ which is the solution of classical Maxwell equations. However, since $\boldsymbol{E}$ is not a canonical variable such an approach leads to a number of fundamental inconsistencies. Thus, Maxwell equations do not follow from the quantum mechanical formalism in a general case of an anisotropic medium. For instance, as was shown in Ref. [50, 51], $\frac{\partial \widehat{\boldsymbol{B}}}{\partial t} = \frac{i}{\hbar}[\widehat{\mathcal{H}}, \widehat{\boldsymbol{B}}] \neq -\nabla \times \widehat{\boldsymbol{E}}$ even in a linear dispersionless limit. This suggests that the quantization of the electromagnetic field has to be modified.

A proper quantum theory starts with identifying canonical variables and formulating a Hamiltonian of the system in terms of these variables. The magnetic vector potential $\boldsymbol{A}$ and scalar electric potential φ are typically chosen as such canonical variables. In an isotropic case quantization is easily formulated with the use of Coulomb gauge conditions, i.e., $\nabla \boldsymbol{A} = 0$ and $\varphi = 0$. However, for an anisotropic system $\nabla \boldsymbol{A} \neq 0$ since $\nabla \boldsymbol{E} \neq 0$ (this is actually one of the main reasons behind commonly employed direct quantization of the electric field in complex systems).

Instead, in the absence of external charges $\nabla \boldsymbol{D} = 0$ is always fulfilled [51]. This suggests the use of the electric vector potential $\boldsymbol{Z}$ ($\boldsymbol{D} = \nabla \times \boldsymbol{Z}$) as a canonical variable for electromagnetic field quantization. Importantly, $\boldsymbol{Z}$ satisfies Coulomb-like condition (i.e., $\nabla \boldsymbol{Z} = 0$ for any medium with $\mu = 1$). In the case of uniaxial crystals considered in this paper, this condition physically originates from the fact that $(\boldsymbol{D}, \boldsymbol{B})$ field is transverse inside the crystal (i.e., $\boldsymbol{D} \perp \boldsymbol{B} \perp \boldsymbol{k}$).

For the sake of completeness, we have used this formulation and developed a corresponding quantum mechanical model for spontaneous parametric down conversion in a general lossy dispersive anisotropic crystal. It is easy to show that the classical Hamiltonian of the system in CGS units can be expressed as:

$$\mathcal{H} = \frac{1}{2} \sum_\omega \int_V d^3r \left[ B^2 + \sum_l \frac{\partial \beta_l^{(1)} \omega}{\partial \omega} D_l^2 \right]$$

$$+ \int_V d^3r \sum_{lmn} \left[ \beta_{lmn}^{(2)} D_l^*(\omega_p) D_m(\omega_s) D_n(\omega_i) + c.c. \right] \quad (F23)$$

where $\boldsymbol{D} = \nabla \times \boldsymbol{Z}$, $\boldsymbol{B} = \frac{1}{c} \frac{\partial \boldsymbol{Z}}{\partial t}$, $\bar{\bar{\beta}}^{(1)}$ and $\bar{\bar{\beta}}^{(2)}$ are first and second order electric permeabilities, $\bar{\bar{\beta}}^{(1)} = \bar{\bar{\varepsilon}}^{-1}$ and $\bar{\bar{\beta}}_l^{(2)} = -\sum_m \beta_{lm}^{(1)}(\omega_s) \bar{\bar{\chi}}_m^{(2)} \bar{\bar{\beta}}^{(1)}(\omega_i) \frac{1}{n_p^2}$.

Next quantizing the electric vector potential $\boldsymbol{Z}$ in a standard way and making derivations similar the ones shown above in this section, we get the expression for the signal photon power emission. We do not present such an analysis here.

In our calculations of the signal photon emission rate and power, we used both formalisms, i.e., the one based on electric field quantization and another one based on the electric vector potential quantization. Both of these theories give similar results.

1. B. Lounis and M. Orrit, *Single-Photon Sources,* Rep. Prog. Phys. **68**, 1129-1179 (2005).
2. M.D. Eisaman, J. Fan, A. Migdall, and S.V. Polyakov, *Invited Review Article: Single-Photon Sources and Detectors,* Rev. Sci. Instrum. **82**, 071101 (2011).
3. P. Lodahl, S. Mahmoodian, and S. Stobbe, *Interfacing Single Photons and Single Quantum Dots With Photonic Nanostructures, Rev. Mod. Phys.* **87**, 347 (2015).
4. P.G. Kwiat, E. Waks, A.G. White, I. Appelbaum, and P.H. Eberhard, *Ultrabright Source of Polarization-Entangled Photons,* Phys. Rev. A **60**, R773(R) (1999).
5. C. Reimer *et al.*, *Cross-Polarized Photon-Pair Generation and Bi-Chromatically Pumped*

*Optical Parametric Oscillation on a Chip,* Nature Commun. **6**, 8236 (2015).
6. M. Fortsch *et al.*, *A Versatile Source of Single Photons for Quantum Information Processing,* Nature Commun. **4**, 1818 (2013).
7. J.W. Silverstone *et al.*, *On-Chip Quantum Interference Between Silicon Photon-Pair Sources,* Nature Photon. **8**, 104-108 (2014).
8. H.J. Kimble, *The Quantum Internet,* Nature **453**, 1023-1030 (2008).
9. A. Aspuru-Guzik and P. Walther, *Photonic Quantum Simulators,* Nature Phys. **8**, 285-291 (2012).
10. L.K. Shalm *et al.*, *Strong Loophole-Free Test of Local Realism*. Phys. Rev. Lett. **115**, 250402 (2015).
11. S. Tanzilli *et al.*, *PPLN Waveguide for Quantum Communication,* Eur. Phys. J. D **18**, 155-160 (2002).
12. M. Fiorentino, S.M. Spillane, R.G. Beausoleil, T.D. Roberts, P. Battle, and M.W. Munro, *Spontaneous Parametric Down-Conversion in Periodically Poled KTP Waveguides and Bulk Crystals,* Opt. Express **15**, 7479-7488 (2007).
13. R. Horn *et al.*, *Monolithic Source of Photon Pairs,* Phys. Rev. Lett. **108**, 153605 (2012).
14. H. Jin e*t al.*, *On-Chip Generation and Manipulation of Entangled Photons Based on Reconfigurable Lithium-Niobate Waveguide Circuits,* Phys. Rev. Lett. **113**, 103601 (2014).
15. D. Grassani *et al.*, *Micrometer-Scale Integrated Silicon Source of Time-Energy Entangled Photons*, Optica **2**, 88 (2015).
16. Q. Li, M. Davanco, and K. Srinivasan, *Efficient and Low-Noise Single-Photon-Level Frequency Conversion Interfaces Using Silicon Nanophotonics*. Nature Photon. **10**, 406-414 (2016).
17. K. Koch, E.C. Cheung, G.T. Moore, S.H. Chakmakjian, and J.M. Liu, *Hot Spots in Parametric Fluorescence with a Pump Beam of Finite Cross Section,* IEEE J. Quant. Electron. **31**, 769 (1995).
18. J. Lee *et al.*, *Giant Nonlinear Response From Plasmonic Metasurfaces Coupled to Intersubband Transitions,* Nature **511**, 65-69 (2014).
19. M.Z. Alam, I. de Leon, and R.W. Boyd, *Large Optical Nonlinearity of Indium Tin Oxide in its Epsilon-Near-Zero Region.* Science **352**, 795-797 (2016).
20. D. de Ceglia, M.A. Vincenti, S. Campione, F. Capolino, J.W. Haus, and M. Scalora, *Second-Harmonic Ddouble-Resonance Cones in Dispersive Hyperbolic Metamaterials*, Phys. Rev. B **89**, 075123 (2014).
21. G. Marino, P. Segovia, A.V. Krasavin, P. Ginzburg, N. Olivier, G.A. Wurtz, and A.V. Zayats, *Second-Harmonic Generation from Hyperbolic Plasmonic Nanorod Metamaterial Slab,* arXiv 1508.07586 (2015).
22. C. Duncan, L. Perret, S. Palomba, M. Lapine, B.T. Kuhlmey, and C.M. de Sterke, *New Avenues for Phase Matching in Nonlinear Hyperbolic Metamaterials,* Sci. Rep. **5**, 8983 (2015).
23. H.N.S. Krishnamoorthy, Z. Jacob, E. Narimanov, I. Kretzschmar, V.M. Menon, *Topological Transitions in Metamaterials,* Science **336**, 205-209 (2012).
24. A.M. Mahmoud and N. Engheta, *Wave-Matter Interactions in Epsilon-and-Mu-Near-Zero Structures,* Nature Comm. **5**, 5638 (2014).
25. R. Sokhoyan and H.A. Atwater, *Cooperative Behavior of Quantum Dipole Emitters Coupled to a Zero-Index Nanoscale Waveguide, arXiv* 1510.07071 (2015).
26. A.N. Poddubny, I.V. Iorsh, and A.A. Sukhorukov, *Generation of Photon-Plasmon*

*Quantum States in Nonlinear Hyperbolic Metamaterials.* Phys. Rev. Lett. **117**, 123901 (2016).
27. C.L. Cortes, W. Newman, S. Molesky, and Z. Jacob, *Quantum Nanophotonics Using Hyperbolic Metamaterials*, J. Opt. **14**, 063001 (2012).
28. V.P. Drachev, V.A. Podolskiy, and A.V. Kildishev, *Hyperbolic Metamaterials: New Physics Behind a Classical Problem,* Opt. Express **21**, 15048-15064 (2013).
29. E.E. Narimanov and A.V. Kildishev, *Metamaterials: Naturally Hyperbolic,* Nature Photon. **9**, 214-216 (2015).
30. M.M. Choy and R.L. Byer, *Accurate Second-Order Susceptibility Measurements of Visible and Infrared Nonlinear Crystals,* Phys. Rev. B **14**, 1693 (1976).
31. H. Vanherzeele and J.D. Bierlein, *Magnitude of the Nonlinear-Optical Coefficients of $KTiOPO_4$.* Optics Lett. **17**, 982-984 (1992).
32. R. Matsushima, N. Tanaka, O. Sugihara, and N. Okamoto, *Low-Temperature Poling of Dye-Doped Polymers for Nonlinear Optical Devices,* Chem. Lett. 200, (2001).
33. A.A. High, *et al.*, *Visible-Frequency Hyperbolic Metasurface*, Nature **522**, 192-196 (2015).
34. A.J. Hoffman *et al.*, *Negative Refraction in Semiconductor Metamaterials.* Nature Mater. **6**, 946 - 950 (2007).
35. Y. Li, S. Kita, P. Munoz, O. Reshef, D.I. Vulis, M. Yin, M. Loncar, and E. Mazur, *On-Chip Zero-Index Metamaterials,* Nature Photon. **9**, 738-742 (2015).
36. P.B. Johnson and R.W. Christy, *Optical Constants of the Noble Metals,* Phys. Rev. B **6**, 4370 (1972).
37. Y. Liu and X. Zhang, *Metamaterials: a new frontier of science and technology,* Chem. Soc. Rev. **40**, 2494–2507 (2011).
38. N. Engheta and R.W. Ziolkowski, *Metamaterials: physics and engineering explorations* (John Wiley & Sons, 2006).
39. A. Poddubny, I. Iorsh, P. Belov, Y. Kivshar, *Hyperbolic metamaterials*, Nature Photon. **7**, 948-957 (2013).
40. A. Yariv and P. Yeh. *Optical Waves in Crystals: Propagation and Control of Laser Radiation.* (Wiley-Interscience, 2002).
41. D.E. Zelmon, D.L. Small, and D. Jundt, *Infrared corrected Sellmeier Coefficients for Congruently Grown Lithium Niobate and 5 mol.% Magnesium Oxide-Doped Lithium Niobate,* J. Opt. Soc. Am. B **14**, 3319-3322 (1997).
42. G.E. Jellison, *Optical Functions of GaAs, GaP, and Ge Determined by Two-Channel Polarization Modulation Ellipsometry*, Opt. Mat. **1**, 151-160 (1992).
43. M.O. Scully and M.S. Zubairy. *Quantum Optics.* 1st ed. (Cambridge University Press, 1997).
44. H.T. Dung, L. Knoll, and D.-G. Welsch, *Three-Dimensional Quantization of the Electromagnetic Field in Dispersive and Absorbing Inhomogeneous Dielectrics*, Phys. Rev. A **57**, 3931 (1998).
45. R. Matloob, R. Loudon, S.M. Barnett, and J. Jeffers, *Electromagnetic Field Quantization in Absorbing Dielectrics*, Phys. Rev. A **52**, 4823 (1995).
46. J.E. Sipe, *Photons in Dispersive Dielectrics,* J. Opt. A: Pure Appl. Opt. **11**, 114006 (2009).
47. N.A.R. Bhat and J.E. Sipe, *Hamiltonian Treatment of the Electromagnetic Field in Dispersive and Absorptive Structured media*, Phys. Rev. A **73**, 063808 (2006).
48. L.G. Suttorp and M. Wubs, *Field Quantization in Inhomogeneous Absorptive Dielectrics,*

Phys. Rev. A **70**, 013816 (2004).
49. B. Huttner and S.M. Barnett, "*Quantization of the Electromagnetic Field in Dielectrics*, Phys. Rev. A **46**, 4306 (1992).
50. S.-T. Ho and P. Kumar, *Quantum Optics in a Dielectric: Macroscopic Electromagnetic-Field and Medium Operators for a Linear Dispersive Lossy Medium  - A Microscopic Derivation of the Operators and Their Commutation Relations*, J. Opt. Soc. Am. B **10**, 1620 (1993).
51. M. Hillery and L.D. Mlodinow, *Quantization of Electrodynamics in Nonlinear Dielectric Media*, Phys. Rev. A  **30**, 1860 (1984).
52. R.L. Byer and S.E. Harris, *Power and Bandwidth of Spontaneous Parametric Emission*, Phys. Rev. **168**, 1064 (1968).
53. M.H. Rubin, D.N. Klyshko, Y.H. Shih, and A.V. Sergienko, *Theory of Two-Photon Entanglement in Type-II Optical Parametric Down-Conversion,* Phys. Rev. A **50**, 5122 (1994).
54. L.D. Landau, L.P. Pitaevskii, and E.M. Lifshitz. *Electrodynamics of Continuous Media, Second Edition: Volume 8 (Course of Theoretical Physics).* 2nd ed. (Butterworth-Heinemann, 1984).
55. J.A. Armstrong, N. Bloembergen, J. Ducuing, and P. S. Pershan *Interactions between Light Waves in a Nonlinear Dielectric*, Phys. Rev. **127**, 1918 (1962).
56. P.S. Pershan, *Nonlinear Optical Properties of Solids: Energy Considerations*, Phys. Rev. **130**, 919 (1963).
57. R.W. Boyd. *Nonlinear Optics.* 3rd ed. (Academic Press, 2008).
58. N.W. Ashcroft and N.D. Mermin. *Solid State Physics*. (Cengage Learning, 1976).
59. A.M. Brown, R. Sundararaman, P. Narang, W. A. Goddard, and H. A. Atwater, *Nonradiative Plasmon Decay and Hot Carrier Dynamics: Effects of Phonons, Surfaces, and Geometry*, ACS Nano **10**, 957-966 (2016).